\documentclass{article}

\usepackage{arxiv}

\usepackage[utf8]{inputenc} 
\usepackage[T1]{fontenc}    
\usepackage{hyperref}       
\usepackage{url}            
\usepackage{booktabs}       
\usepackage{amsfonts}       
\usepackage{nicefrac}       
\usepackage{microtype}      
\usepackage{lipsum}		
\usepackage{graphicx}
\usepackage{natbib}
\usepackage{doi}
\usepackage{amsmath}
\usepackage{arydshln}

\title{A Bayesian Approach to GRAPPA Parallel FMRI Image Reconstruction Increases SNR and Power of Task Detection}

\author{ \href{https://orcid.org/0009-0007-5651-371X}{\includegraphics[scale=0.06]{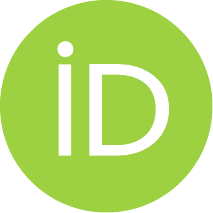}\hspace{1mm}Chase J. Sakitis} \\
        \thanks{\textit{Address for correspondence}: Chase J. Sakitis, Department of Mathematical and Statistical Sciences, Marquette University, Milwaukee, WI, USA. Email: chase.sakitis@marquette.edu}
	Department of Mathematical and Statistical Sciences\\
	Marquette University\\
	Milwaukee, WI 53233 \\
	\texttt{chase.sakitis@marquette.edu} \\
	\And
	\href{https://orcid.org/0000-0002-8713-2158}{\includegraphics[scale=0.06]{orcid.pdf}\hspace{1mm}Daniel B. Rowe} \\
	Department of Mathematical and Statistical Sciences\\
	Marquette University\\
	Milwaukee, WI 53233 \\
	\texttt{daniel.rowe@marquette.edu} \\
}

\renewcommand{\shorttitle}{BGRAPPA}

\begin{document}
\maketitle

\begin{abstract}
In fMRI, capturing brain activation during a task is dependent on how quickly $k$-space arrays are obtained. Acquiring full $k$-space arrays, which are reconstructed into images using the inverse Fourier transform (IFT), that make up volume images can take a considerable amount of scan time. Under-sampling $k$-space reduces the acquisition time, but results in aliased, or “folded,” images. GeneRalized Autocalibrating Partial Parallel Acquisition (GRAPPA) is a parallel imaging technique that yields full images from subsampled arrays of $k$-space. GRAPPA uses localized interpolation weights, which are estimated pre-scan and fixed over time, to fill in the missing spatial frequencies of the subsampled $k$-space. Here, we propose a Bayesian approach to GRAPPA (BGRAPPA) where prior distributions for the unacquired spatial frequencies, localized interpolation weights, and $k$-space measurement uncertainty are assessed from the $a$ $priori$ calibration $k$-space arrays. The prior information is utilized to estimate the missing spatial frequency values from the posterior distribution and reconstruct into full field-of-view images. Our BGRAPPA technique successfully reconstructed both a simulated and experimental single slice image with less artifacts, reduced noise leading to an increased signal-to-noise ratio (SNR), and stronger power of task detection.
\end{abstract}


\section{Introduction}
\subsection{Background}\label{subsec:IntroBackground}
Functional Magnetic resonance imaging (fMRI) is a type of medical imaging developed in the early 1990’s as a technique to noninvasively observe human brain activity without exogenous contrast agents \citep*{bandettini1993fmri}. This procedure examines the brain in action by detecting changes in the brain using the blood-oxygen-level dependent (BOLD) contrast \citep{ogawa1990oxygenation}. The increase in the BOLD contrast in the area of a neuron is a correlate for neuronal firing. Measurements from the machine are arrays of complex-valued spatial frequencies called $k$-space \citep{kumar1975nmr}. These complex-valued $k$-space arrays are then reconstructed into images using a 2D inverse Fourier transform (IFT) producing complex-valued brain images. The magnitude and phase of the complex-valued reconstructed images can be utilized for activation analysis  \citep{rowe2004fmri,rowe2005complex}, but generally only the magnitude is used \citep{bandettini1993fmri}.

In fMRI, measuring full arrays of data for all the slices that form each volume image typically takes about one to two seconds, limiting the temporal resolution of the obtained images and potentially diminishing brain activity detection. Shortening the time it takes to acquire the data required for volume images would improve capturing brain activity. A great deal of work has been dedicated to reducing the acquisition time of the MRI process by accelerating the number of images obtained per unit of time. \citet*{hyde1986parallel}, \citet{pruessmann1999sense}, and \citet*{griswold2002grappa} explored parallel imaging techniques while \citet{lindquist20083Dfmri} measured less three dimensional $k$-space and filtered to expand into the full volume $k$-space. The purpose of each of these techniques was to reduce the acquisition time in MRI.

\subsection{Previous Approach}\label{subsec:IntroPrev}
Historically, a single channel coil receiver has been utilized in fMRI to measure full-sampled $k$-space data arrays. Reducing time between successive volume images is the primary goal of parallel imaging which can also reduce total scan time. More recently, the technological development focus has been to reduce acquisition time by measuring less data without losing the ability to form a full image. This can be achieved by skipping the acquisition of lines in the $k$-space array, i.e. subsampling. To accomplish this, multiple receiver coils are utilized in parallel to obtain spatial frequency arrays which are reconstructed into coil-specific brain images.

\begin{figure}[!h]
	\centering
	\includegraphics[width=5.5in]{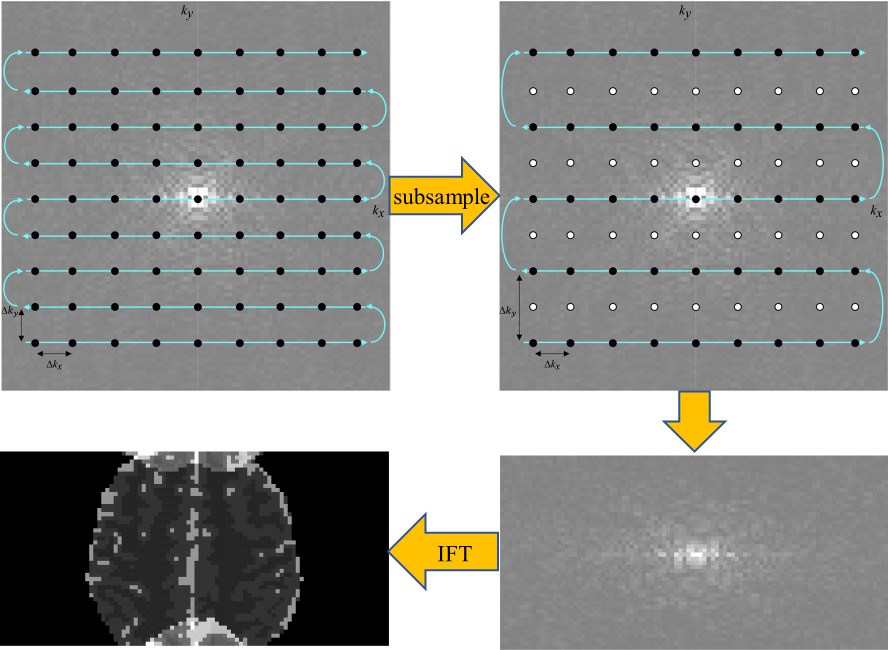}
	\caption{Full $k$-space array (top left), sequence of the subsampled $k$-space array with ${n}_{A}= 2$ (top right), the acquired subsampled $k$-space array (bottom right), and aliased brain image after IFT of the subsampled $k$-space array (bottom left).}
 \label{fig:subsampleimageift}
\end{figure}

Skipping lines in $k$-space introduces what is called an acceleration factor. The acceleration factor indicates which lines of $k$-space data are measured. For example, with an acceleration factor of ${n}_{A}=2$, every other line horizontally in $k$-space is measured. Figure \ref{fig:subsampleimageift} shows the sequential pattern for a fully sampled $k$-space array (top left) compared to a subsampled $k$-space array with an acceleration factor of ${n}_{A}=2$ (top right). This acceleration factor will cause the reconstructed coil images to appear as if they are folded over, because the Fourier transform cannot uniquely map the down sampled signals. We can see an example of this in the bottom left of Figure \ref{fig:subsampleimageift} where the IFT of the subsampled $k$-space causes the brain image to be aliased.

To obtain a full field-of-view (FOV) image, the unacquired spatial frequencies need to be estimated to have full coil $k$-space arrays. The full $k$-space arrays (acquired plus estimation) for each coil are averaged to yield a single, full spatial frequency array. Then, the averaged, full $k$-space array is inverse Fourier transformed into a full brain image. A common method that estimates the unacquired coil spatial frequencies is GeneRalized Autocalibrating Partial Parallel Acquisition (GRAPPA) and was introduced by Griswold et al. (2002). GRAPPA operates in the spatial frequency domain before the IFT, utilizing localized weights to interpolate the missing values in each coil $k$-space array. GRAPPA has its deficiencies, such as low image quality, a low SNR, and diminished task detection power with higher acceleration factors. Bayesian methodologies have been utilized in $k$-space to improve spatial resolution and image quality \citep{kornak2010kbayes}, but here we aim to use it for reconstructing subsampled $k$-space data to produce full brain images. We propose a Bayesian approach to GRAPPA that will incorporate prior information, yielding increased SNR and image quality, with improved task detection power.

\subsection{Overview}\label{subsec:IntroOverview}
The second section of this paper will explain the model of GRAPPA image reconstruction and formulate the complex-valued problem as a real-valued isomorphic representation. This will lead into our proposed Bayesian approach presented in Section \ref{sec:BGRAPPA}. Section \ref{sec:SimData} will show results from comparing traditional GRAPPA and our new BGRAPPA approach to simulated non-task and task fMRI data. Section \ref{sec:ExpData} presents a similar comparison with experimental task fMRI data. We will conclude in Section \ref{sec:Discussion} with an overview of the important results of the paper and a discussion of future work.

\section{GRAPPA Technique}\label{sec:GRAPPA}
\subsection{Reconstruction Process}\label{subsec:Problem}
\begin{figure}[!b]
	\centering
	\includegraphics[width=5.6in]{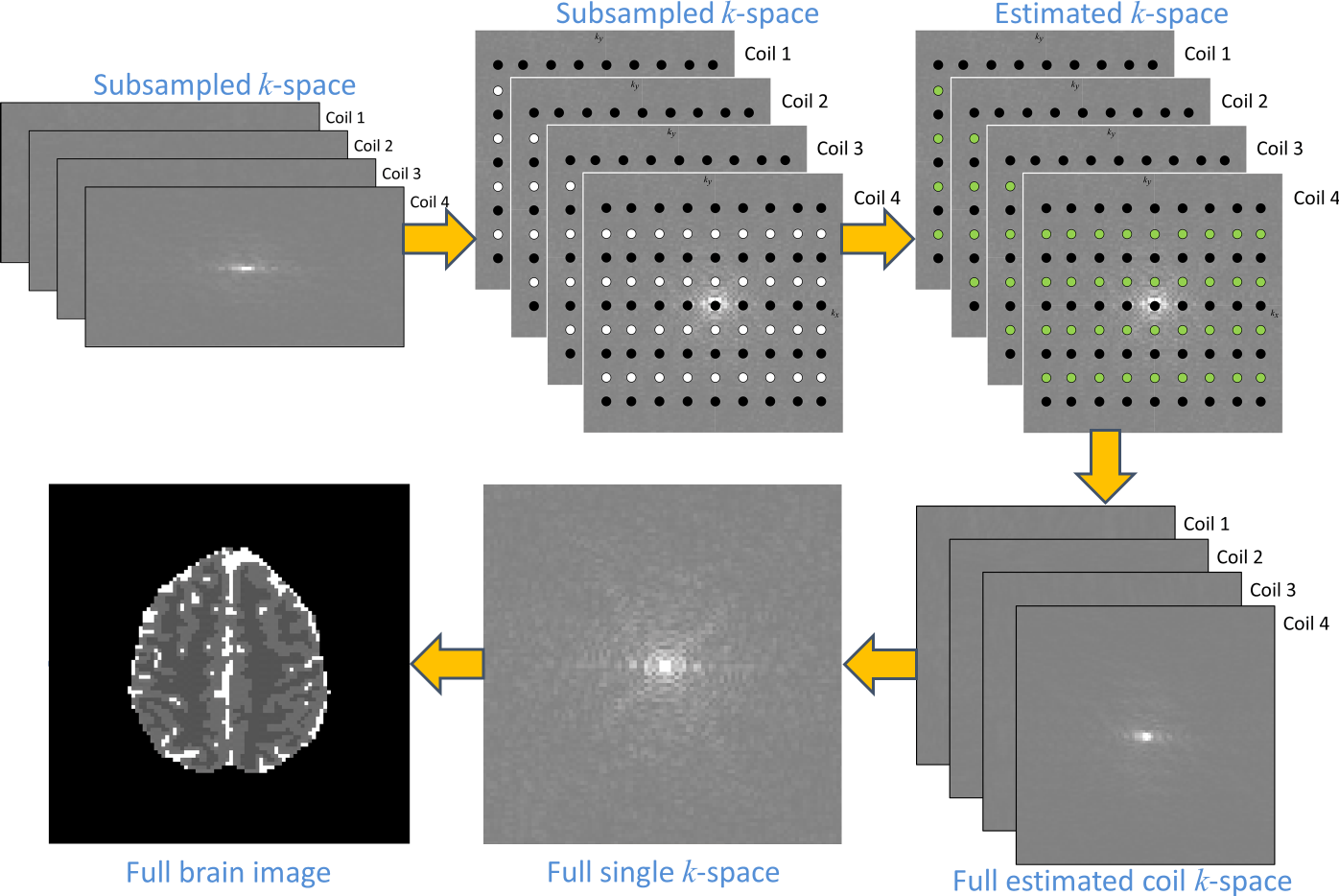}
	\caption{Subsampled $k$-space coil arrays (top left) that are spread out to show a full $k$-space array where the black dots are the acquired spatial frequencies, and the white dots are the unacquired spatial frequencies (top middle). The missing spatial frequencies are then estimated (green dots in the top right) yielding full coil $k$-space arrays (bottom right). The full coil $k$-space arrays are averaged together to produce a full spatial frequency array (bottom middle) which is then transformed into a full brain image (bottom left) using the IFT.}
 \label{fig:grappaprocess}
\end{figure}
As mentioned in Subsection 1.2, to measure less $k$-space data and still produce a full brain image, ${n}_{C}>1$  receiver coils must be utilized. The process for GRAPPA is exhibited in Figure \ref{fig:grappaprocess} with an illustrative example of using ${n}_{C}=4$  coils. The machine acquires subsampled spatial frequency arrays for each of the four coils shown in the top left of Figure \ref{fig:grappaprocess}. The top middle of Figure \ref{fig:grappaprocess} displays the subsampled $k$-space arrays as full arrays with the black dots indicating the acquired spatial frequencies and the white dots indicating the unacquired spatial frequencies. The unacquired spatial frequencies are estimated using GRAPPA image reconstruction, displayed as the green dots in the top right of Figure \ref{fig:grappaprocess}. This yields full coil $k$-space arrays as shown in the bottom right of Figure \ref{fig:grappaprocess}. To get a single full spatial frequency array (bottom middle), the full coil spatial frequency arrays are averaged together. The full spatial frequency is then IFT reconstructed into a single, full FOV brain image (bottom left of Figure \ref{fig:grappaprocess}).

\subsection{Model}\label{subsec:GRAPPAModel}
\begin{figure}[!b]
	\centering
	\includegraphics[width=5.6in]{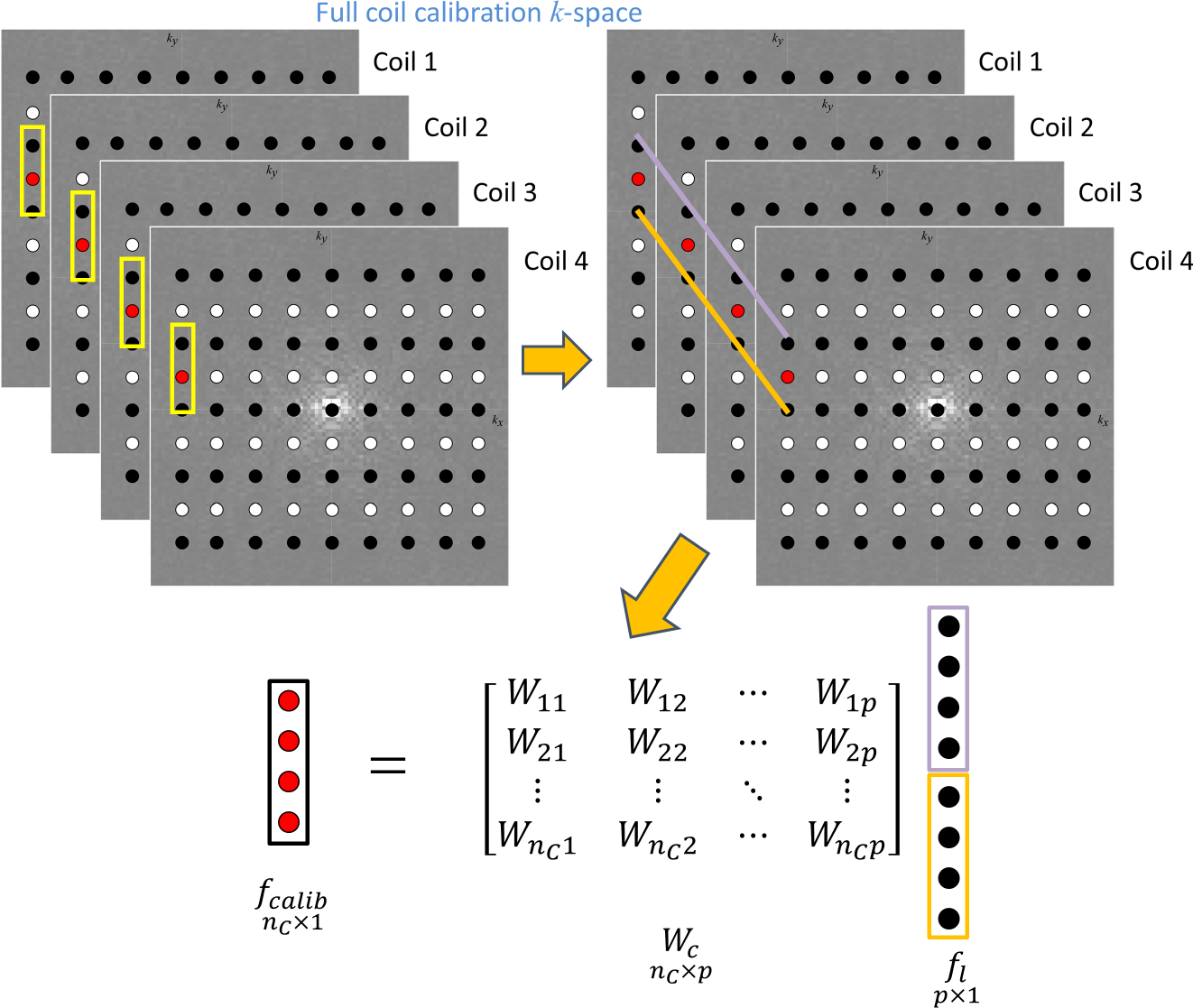}
	\caption{The $k$-space coil arrays in the top left are fully sampled where the black dots are treated as the acquired spatial frequencies and the red dots are the calibration points utilized to calculate the weights for those coil spatial frequencies. The yellow box shows a 2$\times$1 kernel indicating which points are utilized to estimate the weights. From this, we get an acquired black dot above and below each red coil calibration point. The black points above the calibration points are then stacked by coil (1 through 4) which is then placed above the stacked black dots below the calibration points. The image in the top right demonstrates the stacking of the black dots through the purple line and the orange line with the purple vector then being placed above the orange vector in the system of linear equations (bottom). The weights, ${W}_{c}$, for those “unacquired” coil spatial frequencies are then estimated using least squares. Once the weights have been estimated, the red calibration points move to the next white dots to estimate the set of weights for the next unacquired spatial frequencies.}
 \label{fig:kernelweights}
\end{figure}
In GRAPPA, the complex-valued localized interpolation weights are estimated using pre-scan coil calibration spatial frequency arrays. These coil calibration $k$-space arrays are fully sampled coil spatial frequency arrays that are collected prior to the actual fMRI experiment. Kernels of varying sizes can be used to estimate the weights, creating a system of linear equations. Figure \ref{fig:kernelweights} illustrates how a 2×1 kernel is utilized to estimate the weights from the full coil calibration spatial frequencies with a four-channel coil array. In Figure \ref{fig:kernelweights}, all the complex-valued data points are acquired, but are treated differently depending on the location of the data point. The black data points, ${f}_{l}$, are utilized as the “acquired” complex-valued spatial frequency values, the red points, ${f}_{calib}$, are the complex-valued calibration spatial frequency points, and the white points are ignored for the calculation of those weights associated with the current ${f}_{calib}$ points. The white dots represent the spatial frequencies that would be unacquired during the fMRI experiment but are used as calibrations points to estimate the complex-valued weights for those spatial frequencies.

The calibration points ${f}_{calib}$ and the “acquired” spatial frequencies ${f}_{l}$ along with the unacquired complex-valued weights, ${W}_{c}$, create a system of linear equations as displayed in Figure 3 (bottom). From the linear equations, we can estimate the weights ${w}_{c}$ using Eq. \ref{eqn:sysofeqabbr},
\begin{equation}\label{eqn:sysofeqabbr}\tag{2.1}
{{W}_{c}}^{(\nu)}={{{f}_{calib}}^{(\nu)}}{{{f}_{l}}^{(\nu)}}^{\dag}({{{f}_{l}}^{(\nu)}}{{{f}_{l}}^{(\nu)}}^{\dag})^{-1},\hspace{4mm}\nu=1,...,K
\end{equation}
where ${W}_{c}\in\mathbb{C}^{{n}_{C}\times{p}}$ is the complex-valued weights, ${f}_{calib}\in\mathbb{C}^{{n}_{C}\times1}$ is the complex-valued calibration spatial frequencies, ${f}_{l}\in\mathbb{C}^{{p}\times1}$ is the “acquired” complex-valued spatial frequencies, $p={n}_{c}{k}_{rows}{k}_{cols}$, ${k}_{rows}$ is the number of rows in the kernel, ${k}_{cols}$ is the number of columns in the kernel, $\dag$ is the Hermitian or conjugate transpose, and $K$ is the total number of unacquired spatial frequencies in the subsampled $k$-space array. The process is repeated for each spatial frequency point that would be unacquired during the actual fMRI experiment (the white dots in Figure \ref{fig:kernelweights}), yielding different weights for each unacquired spatial frequency.

Once the weights for each of the unacquired coil spatial frequencies are estimated from the calibration $k$-space arrays, those weights are then utilized to interpolate the unacquired spatial frequencies in the actual fMRI experiment. The GRAPPA model with the estimated weights becomes
\begin{equation}\label{eqn:complexmodel}\tag{2.2}
{{f}_{ec}}^{(\nu)}={{W}_{c}}^{(\nu)}{{f}_{kc}}^{(\nu)}+{{\eta}_{c}}^{(\nu)},\hspace{10mm}\nu=1,...,K
\end{equation}
where ${f}_{ec}\in\mathbb{C}^{{n}_{C}\times1}$ is the complex-valued interpolated $k$-space values, ${f}_{kc}\in\mathbb{C}^{{p}\times1}$ is the complex-valued acquired $k$-space values, and ${\eta}_{c}\in\mathbb{C}^{{n}_{C}\times1}$ is the additive complex-valued noise with ${\eta}_{c}\sim N(0, {\tau}^{2}(1+i))$. The interpolated coil $k$-space values, ${f}_{ec}$, are inserted in the respective locations of each coil yielding full coil $k$-space arrays (top right of Figure \ref{fig:grappaprocess}).

With GRAPPA image reconstruction, however, the resulting reconstructed brain images can have diminished SNR which is a consequence of either a decreased signal intensity, increased temporal noise variance, or a combination of the two. With an increase in the temporal noise variance, this can lead to reduced power in task detection as well. These deficiencies motivate our Bayesian approach, which will allow for a more automated method for image reconstruction without having to potentially store and use large matrices. Unlike GRAPPA, our Bayesian approach will utilize all valuable available prior information from the calibration spatial frequency arrays and provide full distributions for the unacquired spatial frequencies, the weights, and the residual $k$-space variance.

\subsection{GRAPPA Isomorphic Representation}\label{sec:GRAPPAIso}
Since the model is complex-valued, we can write the GRAPPA model in a real-valued isomorphic representation. The traditional GRAPPA model estimates the unacquired spatial frequencies while the data values are still in complex-valued form shown in Eq \ref{eqn:GRAPPAcomplexnumbers}.
\begin{equation}\label{eqn:GRAPPAcomplexnumbers}\tag{2.3}
({f}_{eR}+i{f}_{eI})=({W}_{R}+i{W}_{I})({f}_{kR}+i{f}_{kI})+({\eta}_{R}+i{\eta}_{I}).
\end{equation}
This complex-valued model can be expressed by a real-valued isomorphic representation as conveyed by Eq. \ref{eqn:sysofeqGRAPPAIso}.
\begin{equation}\label{eqn:sysofeqGRAPPAIso}\tag{2.4}
\left[ \begin{array}{c}
{f}_{eR}\\ \hdashline[2pt/2pt]
{f}_{eI}
\end{array} \right]
\hspace{-1mm}=\hspace{-1mm}
\left[ \begin{array}{c;{2pt/2pt}c}
{W}_{R} & -{W}_{I}\\ \hdashline[2pt/2pt]
{W}_{I} & {W}_{R}
\end{array} \right]
\left[ \begin{array}{c}
{f}_{kR}\\ \hdashline[2pt/2pt]
{f}_{kI}
\end{array} \right]
\hspace{-1mm}+\hspace{-1mm}
\left[ \begin{array}{c}
{\eta}_{R}\\ \hdashline[2pt/2pt]
{\eta}_{I}
\end{array} \right],
\hspace{6mm}({\eta}_{R},{\eta}_{I})'\sim N(0,{\tau}^{2}{I}_{2{n}_{C}}).
\end{equation}
Eq. \ref{eqn:sysofeqGRAPPAIso} characterizes the design matrix $W$ as being skew-symmetric. The proposed BGRAPPA model will use the real-valued isomorphism (Eq. \ref{eqn:sysofeqGRAPPAIso}) instead of the complex-valued representation (Eq. \ref{eqn:complexmodel}).

\section{Bayesian Approach to GRAPPA}\label{sec:BGRAPPA}
For our proposed Bayesian approach, we use the same linear model as GRAPPA as expressed Eq. \ref{eqn:complexmodel}, except the acquired spatial frequencies will be the ${f}_{ec}$ variable instead of the ${f}_{kc}$ variable. This creates a model where the design matrix and the coefficients can both be treated as unknown parameters, allowing us to take a Bayesian approach to the linear regression. Then the weights, ${W}_{c}$, and the unacquired spatial frequencies, ${f}_{kc}$, along with the residual $k$-space variance, ${\tau}^{2}$, are treated as unknowns with prior distributions placed on them. We also use an isomorphic real-valued representation of the linear GRAPPA model in Eq. \ref{eqn:complexmodel} and is given by
\begin{equation}\label{eqn:sysofeq}\tag{3.1}
\begin{bmatrix}
{f}_{eR}\\ {f}_{eI}
\end{bmatrix}
=
\begin{bmatrix}
{W}_{R} & -{W}_{I}\\
{W}_{I} & {W}_{R}
\end{bmatrix}
\begin{bmatrix}
{f}_{kR}\\ {f}_{kI}
\end{bmatrix}+
\begin{bmatrix}
{\eta}_{R}\\ {\eta}_{I}
\end{bmatrix},
\end{equation}
where ${f}_{eR}\in\mathbb{R}^{{n}_{C}\times1}$ and ${f}_{eI}\in\mathbb{R}^{{n}_{C}\times1}$ are the real and imaginary components, respectively, of ${f}_{ec}$, ${W}_{R}\in\mathbb{R}^{{n}_{C}\times{p}}$ and ${W}_{I}\in\mathbb{R}^{{n}_{C}\times{p}}$ are the real and imaginary components of ${W}_{c}$, ${f}_{kR}\in\mathbb{R}^{{p}\times1}$ and ${f}_{kI}\in\mathbb{R}^{{p}\times1}$ are the real and imaginary components of ${f}_{kc}$, and ${\eta}_{R}\in\mathbb{R}^{{n}_{C}\times1}$ and ${\eta}_{I}\in\mathbb{R}^{{n}_{C}\times1}$ are the real and imaginary components of ${\eta}_{c}$ with $({\eta}_{R},{\eta}_{I})'\sim N(0,{\tau}^{2}{I}_{{2n}_{C}})$. This equation is a latent variable model with complex values and can be more compactly written as ${f}_{e}=W{f}_{k}+\eta$ where ${f}_{e}\in\mathbb{R}^{2{n}_{C}\times1}$, $W\in\mathbb{R}^{2{n}_{C}\times2p}$, ${f}_{k}\in\mathbb{R}^{2p\times1}$, and $\eta\in\mathbb{R}^{2{n}_{C}\times1}$ are the real-valued isomorphic representations of ${f}_{ec}$, ${W}_{c}$, ${f}_{kc}$, and ${\eta}_{c}$, respectively.

In this method, two different representations of the weights will be used. The first representation is the proper skew-symmetric design matrix $W$ as shown in Eq. \ref{eqn:sysofeq}. The second representation is $D=[{W}_{R},\hspace{2mm}{W}_{I}]$ which is used in the prior distribution and for parameter estimation of the weights. This is to ensure ${W}_{R}$ and ${W}_{I}$ are uniquely estimated for $W$ and do not need to be duplicated.

\subsection{Data Likelihood, Prior, and Posterior Distributions}\label{subsec:BGRAPPAdistr}
Like GRAPPA, we assume that the residual spatial frequency error is normally distributed in the real and imaginary components, since the real and imaginary components of fMRI data are assumed to be normally distributed \citep{lindquist2008fmri}. The data likelihood for the acquired spatial frequencies for the ${n}_{c}$ coils is
\begin{equation}\label{eqn:likelihood}\tag{3.2}
P({f}_{e}|W,{f}_{k},\tau^{2})\propto(\tau^{2})^{-\frac{2{n}_{C}}{2}}\exp\left[-\frac{1}{2\tau^{2}}({f}_{e}-W{f}_{k})'({f}_{e}-W{f}_{k})\right].
\end{equation}
We can quantify available prior information about the unacquired spatial frequencies ${f}_{k}$, the weights $W$, and the residual $k$-space variance $\tau^{2}$ with assessed hyperparameters of prior distributions. The unacquired spatial frequencies ${f}_{k}$ are specified to have a normal prior distribution, expressed in Eq. \ref{eqn:fkprior}. The weights $D$ are also specified to have a normal prior distribution (Eq. \ref{eqn:Wprior}) and the $k$-space noise variance $\tau^{2}$ is specified to have an inverse gamma prior distribution (Eq. \ref{eqn:tauprior}),
\begin{equation}\label{eqn:fkprior}\tag{3.3}
P({f}_{k}|{n}_{k},{f}_{k0},\tau^{2})\propto(\tau^{2})^{\frac{-2p}{2}} \exp\left[-\frac{{n}_{k}}{2\tau^{2}}({f}_{k}-{f}_{k0})'({f}_{k}-{f}_{k0})\right],
\end{equation}
\begin{equation}
\label{eqn:Wprior}\tag{3.4} 
P(D|{n}_{w},{D}_{0},\sigma^{2})\propto(\tau^{2})^{\frac{-2{n}_{C}p}{2}} \exp\left[-\frac{{n}_{w}}{2\tau^{2}}tr(D-{D}_{0})(D-{D}_{0})'\right],
\end{equation}
\begin{equation}\label{eqn:tauprior}\tag{3.5}
P(\tau^{2}|{\alpha}_{k},\delta)\propto(\tau^{2})^{-({\alpha}_{k}+1)}\exp\left[-\frac{\delta}{\tau^{2}}\right],
\end{equation}
where $tr$ is the trace of the $(D-{D}_{0})(D-{D}_{0})'$ matrix and the hyperparameters ${n}_{k}$, ${f}_{k0}$, ${n}_{w}$, ${D}_{0}$, ${\alpha}_{k}$, and $\delta$ are assessed from the pre-scan calibration spatial frequencies. The joint posterior distribution of the unacquired spatial frequencies ${f}_{k}$, the weights $W$, and the residual $k$-space variance ${\tau}^{2}$ is
\begin{equation}\label{eqn:posterior}\tag{3.6}
P(D,{f}_{k},\tau^{2}|{f}_{e})\propto P({f}_{e}|W,{f}_{k},\tau^{2})P({f}_{k}|{n}_{k},{f}_{k0},\tau^{2}) P(D|{n}_{w},{D}_{0},\tau^{2})P(\tau^{2}|{\alpha}_{k},\delta),
\end{equation}
with the distributions specified from Equations \ref{eqn:likelihood}, \ref{eqn:fkprior}, \ref{eqn:Wprior}, and \ref{eqn:tauprior}.

A technique that can be utilized for parameter estimation is using Markov chain Monte Carlo (MCMC) Gibbs sampling. The Gibbs sampler uses the posterior conditionals to generate the entire distribution for each parameter at each time point yielding more information that can be used for statistical analysis. However, the computation time is longer compared to using an iterative maximum $a$ $posteriori$ (MAP) method. For this paper, we only use the MAP estimate since we would only be interested in the mean of the distributions for each parameter which is theoretically equal to the mode for the unacquired spatial frequencies ${f}_{k0}$ and the weights $W$. 

\subsection{Hyperparameter Determination} \label{subsec:BSENSEhyperparameter}
The hyperparameters can be appropriately assessed in an automated way using the full pre-scan coil calibration spatial frequencies. For the BGRAPPA hyperparameter assessment, the same full calibration spatial frequencies and ${f}_{calib}=W{f}_{l}$ model are used like in GRAPPA reconstruction, but each spatial frequency point is treated differently than GRAPPA. As shown in Figure 4, the calibration spatial frequencies ${f}_{calib}$ for BGRAPPA are in the location of the data points where the acquired spatial frequencies are in the actual fMRI experiment. For GRAPPA, these data points are assigned to the ${f}_{l}$ variable in the ${f}_{calib}={W}_{c}{f}_{l}$ model shown at the bottom of Figure 4. Using Eq. \ref{eqn:sysofeqabbr}, this will result in the prior for the weights in BGRAPPA, ${D}_{0}$, to be different than the estimated weights utilized in GRAPPA image reconstruction. The ${f}_{l}$ points used for estimating the prior mean for the weights are averaged to obtain the prior mean of the unacquired spatial frequencies, ${f}_{k0}$.
\begin{figure}[!t]
	\centering
	\includegraphics[width=5.5in]{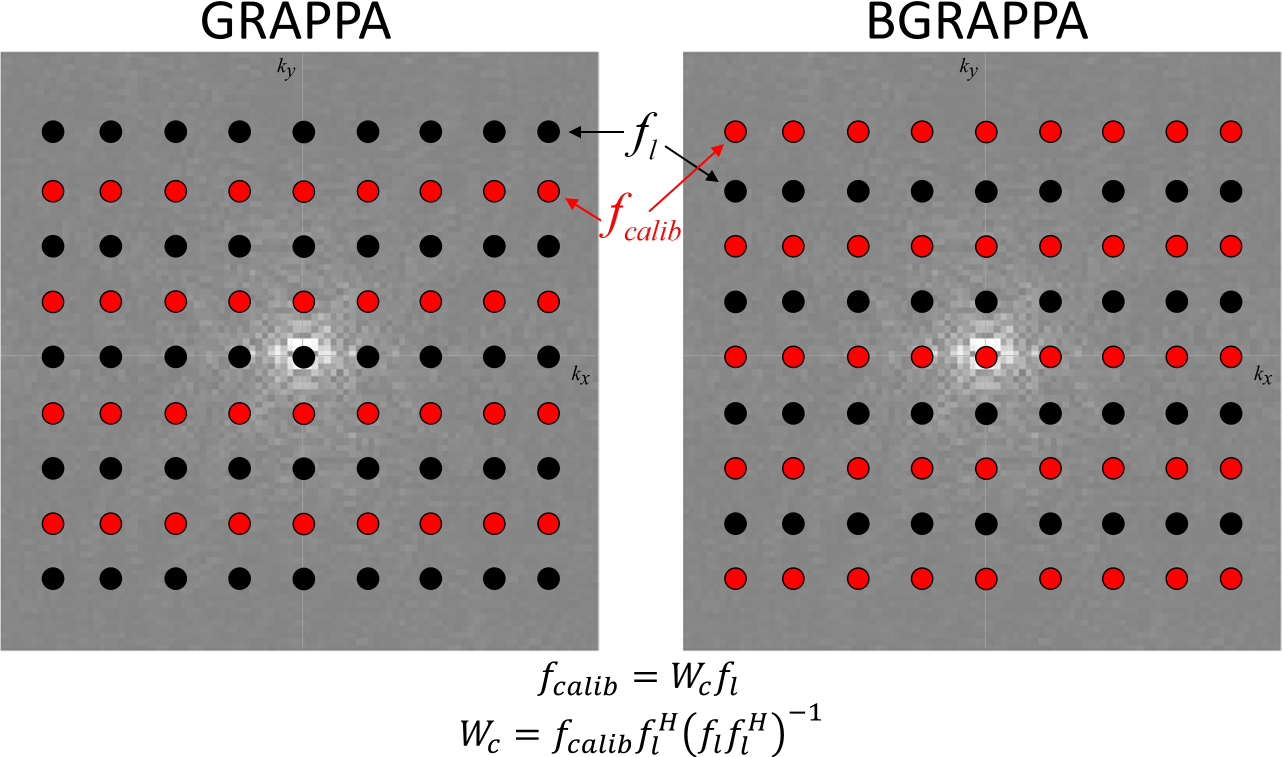}
	\caption{Full calibration $k$-space arrays that indicate which data points are used as ${f}_{calib}$ points and the ${f}_{l}$ points for GRAPPA (left) and BGRAPPA (right).}
 \label{fig:calibrationpoints}
\end{figure} 

The hyperparameters ${n}_{k}$ and ${n}_{w}$, which are the prior scalars of the prior means, are assessed to be the number of calibration time points ${n}_{cal}$. The average residual $k$-space variance over the coil spatial frequency arrays is calculated to obtain a prior mean for the residual $k$-space variance ${\tau}_{0}^{2}$. The hyperparameters ${\alpha}_{k}$ (shape parameter of the inverse gamma) and $\delta$ (scale parameter of the inverse gamma) are assessed to be ${\alpha}_{k}={n}_{cal}-1$ and $\delta=({n}_{cal}-1){\tau}_{0}^{2}$. This prior information is incorporated to estimate the unacquired spatial frequencies in the subsampled $k$-space arrays.

\subsection{Posterior Estimation}\label{subsec:BGRAPPAEstimation}
Using the posterior distribution in Eq. \ref{eqn:posterior}, the MAP estimate for the unacquired spatial frequencies ${f}_{k}$, the weights $W$, and the residual $k$-space variance ${\tau}^{2}$ is estimated via the Iterated Conditional Modes (ICM) optimization algorithm \citep{lindley1972bayes,ohagan1994stat}. Beginning with the prior means for each parameter as initial estimates, the ICM algorithm iterates over the parameters, calculating its posterior conditional mode until convergence at the joint posterior mode. The posterior conditional modes are
\begin{equation}\label{eqn:fkICM}\tag{3.7}
\hat{f}_{k}=(W'W+{n}_{k}{I}_{2p})^{-1}(W'{f}_{e}+{n}_{k}{f}_{k0}),
\end{equation}
\begin{equation}\label{eqn:WICM}\tag{3.8}
\hat{D}=({F}_{e}{F}_{k}'+{n}_{w}{D}_{0})({F}_{k}{F}_{k}'+{n}_{w}{I}_{2p})^{-1},
\end{equation}
\begin{equation}\label{eqn:tauICM}\tag{3.9}
\hat{\tau}^{2}=\frac{\Theta}{2(2{n}_{C}+2p+2{n}_{C}p+1)},
\end{equation}
where $\Theta=({f}_{e}-W{f}_{k})'({f}_{e}-W{f}_{k})+{n}_{k}({f}_{k}-{f}_{k0})'({f}_{k}-{f}_{k0})+{\alpha}_{k}\delta+{n}_{w}tr[(D-{D}_{0})(D-{D}_{0})']$, ${F}_{e}=\left[{f}_{eR},\hspace{1mm}{f}_{eI}\right]$ and ${F}_{k}\in\mathbb{R}^{2p\times2}$ is a skew symmetric matrix representation of the unaliased voxel values ${f}_{k}$ as expressed by
\begin{equation}\label{eqn:Vmat}\tag{3.10}
{F}_{k}=
\begin{bmatrix}
{f}_{kR} & {f}_{kI}\\
-{f}_{kI} & {f}_{kR}
\end{bmatrix}.
\end{equation}

\section{Simulation Study}\label{sec:SimData}
\subsection{Non-Task Spatial Frequency Data}\label{subsec:SimDataGen}
\begin{figure}[!b]
	\centering
	\includegraphics[width=5.5in]{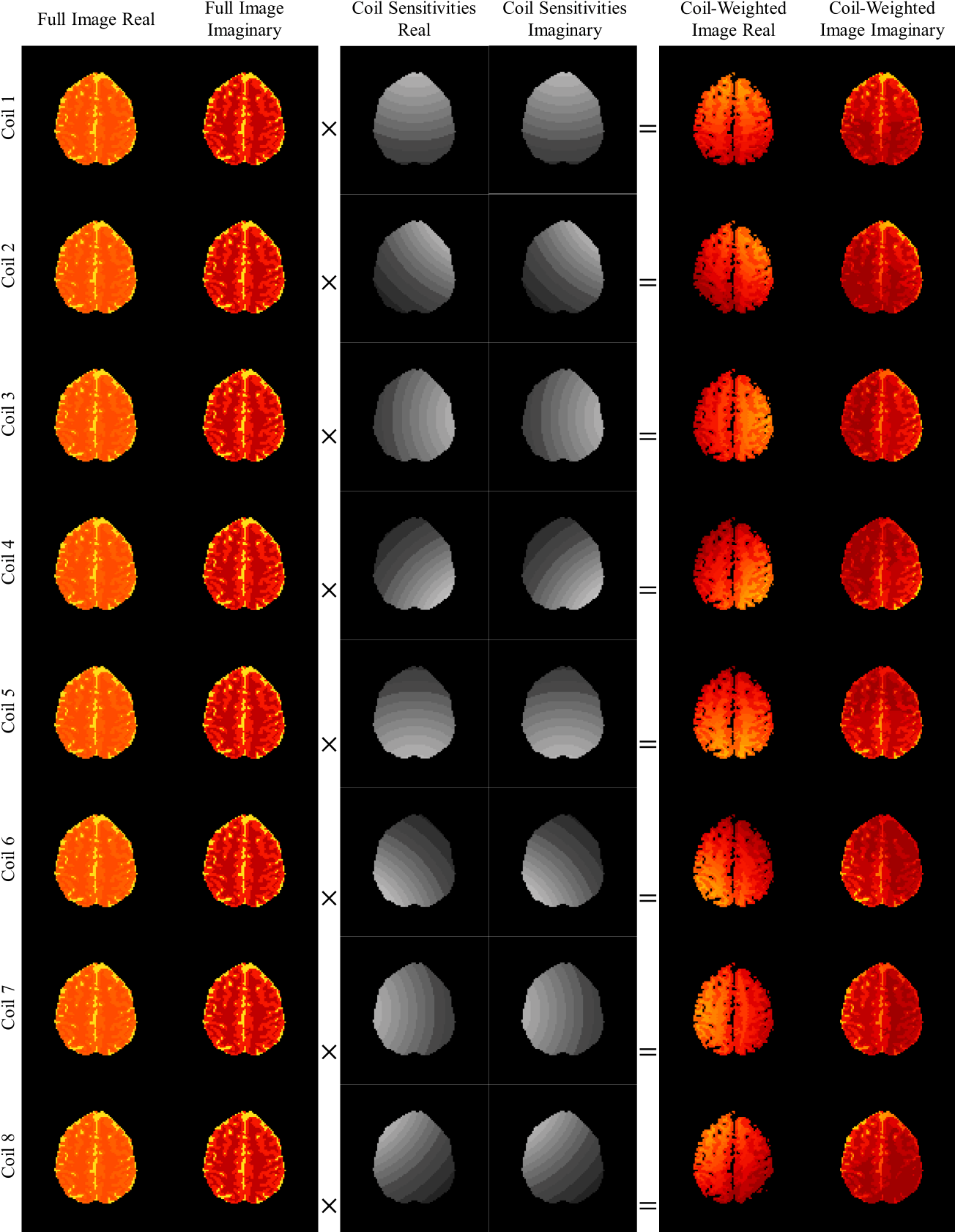}
	\caption{Real and imaginary components of the true complex-valued simulated image (first and second column) voxel-wise multiplied by the real and imaginary parts of the complex-valued coil sensitivities for each of the ${n}_{C}=8$ coils (third and fourth column) yielding the real and imaginary components of the complex-valued coil-weighted images (fifth and sixth column column).}
 \label{fig:weightedcoils}
\end{figure}

A noiseless non-task image was used to create two separate series of ${n}_{TR}=510$ simulated full coil images for one slice to mimic real-world MRI experimental data. The last ${n}_{cal}$ time points of the first time series will serve as the calibration information utilized for hyperparameter assessment. The second time series was used for simulating a subsampled non-task experiment. The complex-valued non-task images were multiplied by a complex-valued designed sensitivity map with ${n}_{C}=8$ coils. Figure \ref{fig:weightedcoils} illustrates the real and imaginary parts of the full simulated brain image (first and second column) being voxel-wise multiplied by the real and imaginary components of the sensitivities for each of the ${n}_{C}=8$ coil (third and fourth column). This results in the real and imaginary components of the complex-valued full coil-weighted images (fifth and sixth column).

In real-world MRI experiments, the first few images in an fMRI time series have increased signal as the magnetization reaches a stable state. To mimic this, the first three of both non-task time series of ${n}_{TR}=510$ time points of the simulated non-task time series were scaled with the signal slightly decreasing from the first to the third time point before reaching a stable signal in the fourth time point. The scaling was determined by dividing the first three images of the experimental data by the 21st image, separately. After dividing the three images, the signal increase for each tissue type (white matter, grey matter and CSF) was averaged together for each of the three divided images, calculating the average signal increase for each matter type. For example, the average signal increase in the first image for the white matter was 40\%, 55\% for the grey matter and 75\% for the CSF giving multiplication factors of 1.40, 1.55, and 1.75 for the matter types, respectively. This process was repeated for the second and third image in the series with the multiplication factors decreasing from the first to the third image.

\begin{figure}[!b]
	\centering
	\includegraphics[width=4.5in]{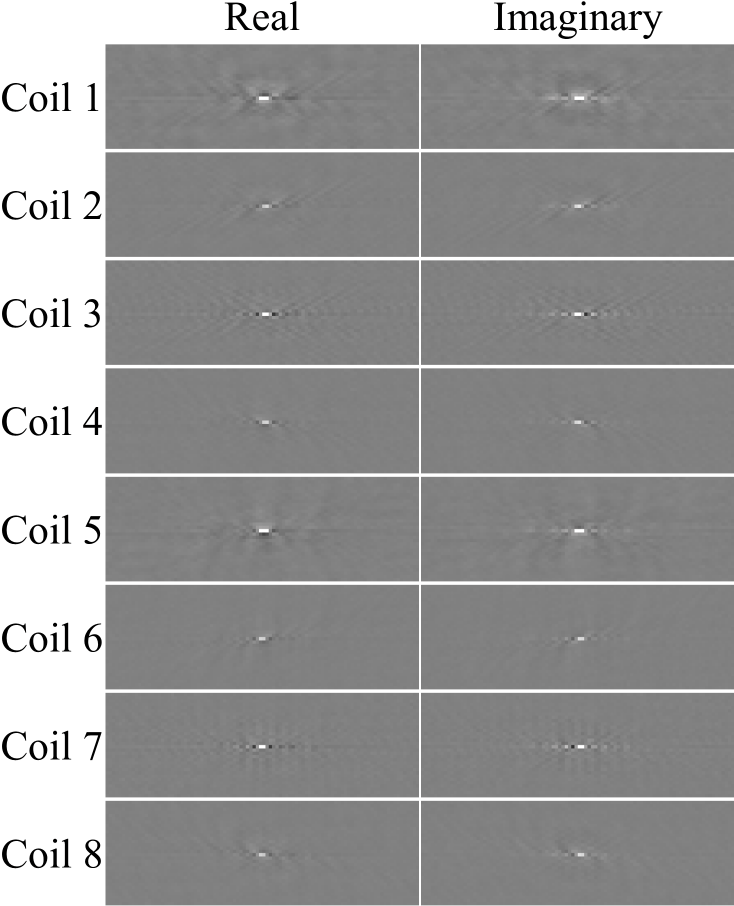}
	\caption{Simulated acquired noisy subsampled coil spatial frequency arrays for the first time point in the non-task time series with an acceleration factor of ${n}_{A}=3$.}
 \label{fig:subsamplecoilmeasurements}
\end{figure}

The series of images for both the ${n}_{cal}$ calibration images and the full simulated images were then Fourier transformed into noiseless full coil $k$-space arrays. The time series of coil $k$-space arrays were simulated by adding separate $N(0,0.0036{n}_{y}{n}_{x})$ noise, where ${n}_{y}$ and ${n}_{x}$ are the number of rows and columns, respectively, in the full $k$-space array, to the real and imaginary parts of full coil $k$-space arrays, corresponding to the noise in the real-world fMRI experimental data. To mimic the fMRI experiment, the first 20 time points of the second time series were omitted leaving 490 time points of spatial frequency arrays for the single slice. However, the first 10 time points of an fMRI experiment can be used to estimate a ${T}_{1}$ map which efficiently segments the different tissue types. The next 10 time points can be utilized to estimate a static magnetic field map to adjust for geometric distortions \citep{karaman2015relaxivities}. The remaining 490 time points in the time series were subsampled by censoring lines in $k$-space according to an acceleration factor of ${n}_{A}=3$. An example of the real and imaginary components of subsampled $k$-space arrays for ${n}_{C}=8$ coils and an acceleration factor of ${n}_{A}=3$ at one time point is exhibited in Figure \ref{fig:subsamplecoilmeasurements}.

\subsection{Non-Task Reconstruction Results}\label{subsec:SimDataResults}
\begin{figure}[!h]
	\centering
	\includegraphics[width=6.4in]{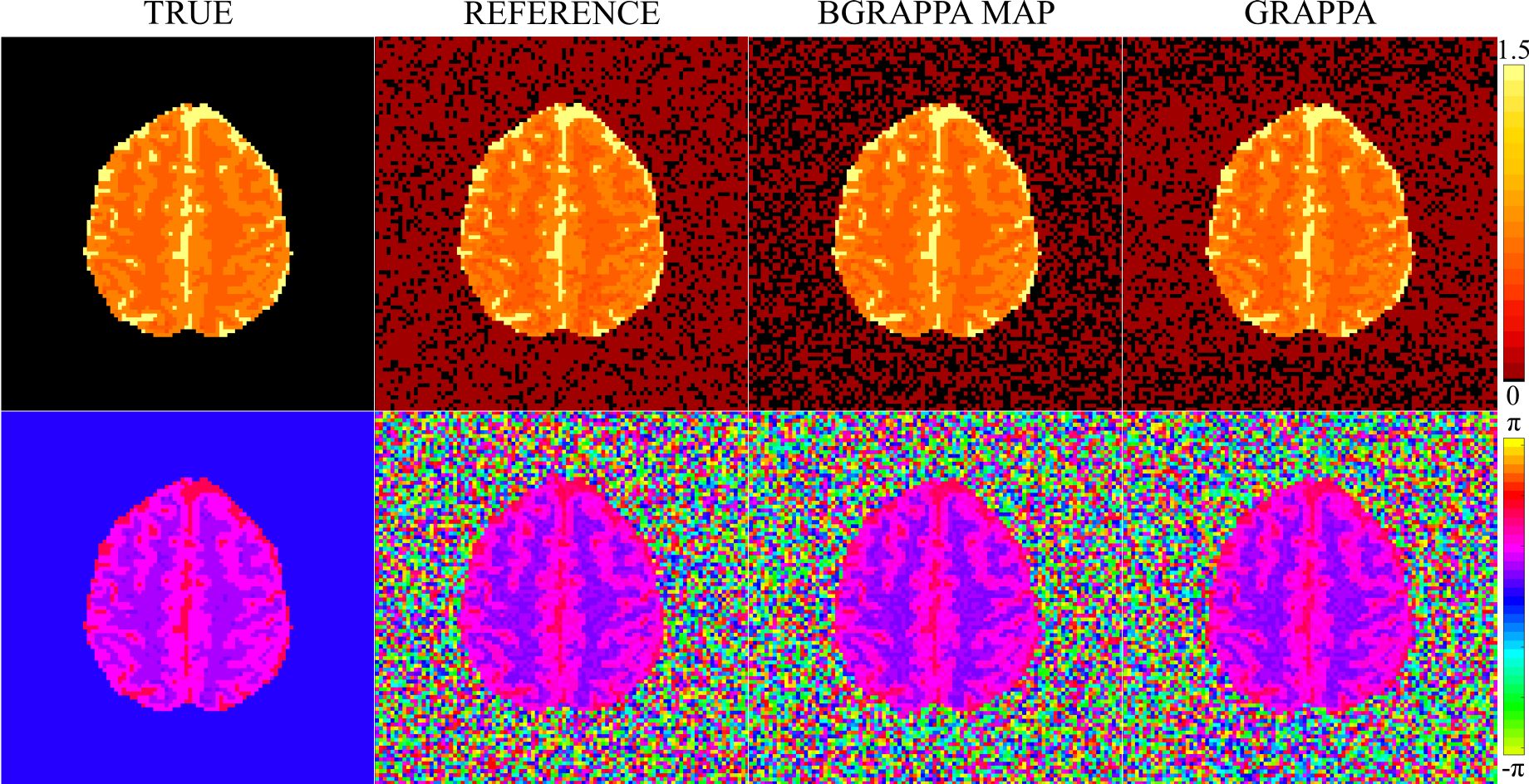}
	\caption{True non-task unaliased images (first column), reference non-task reconstructed images (second column), BGRAPPA MAP unaliased non-task images (third column) using ICM and GRAPPA non-task images (fourth column) with magnitude images in the first row and phase images in the second row. Due to the circular nature of phase angles, the color bar for the phase images have wrap-around.}
 \label{fig:RecSingleTP}
\end{figure}

\begin{figure}[!h]
	\centering
	\includegraphics[width=6.4in]{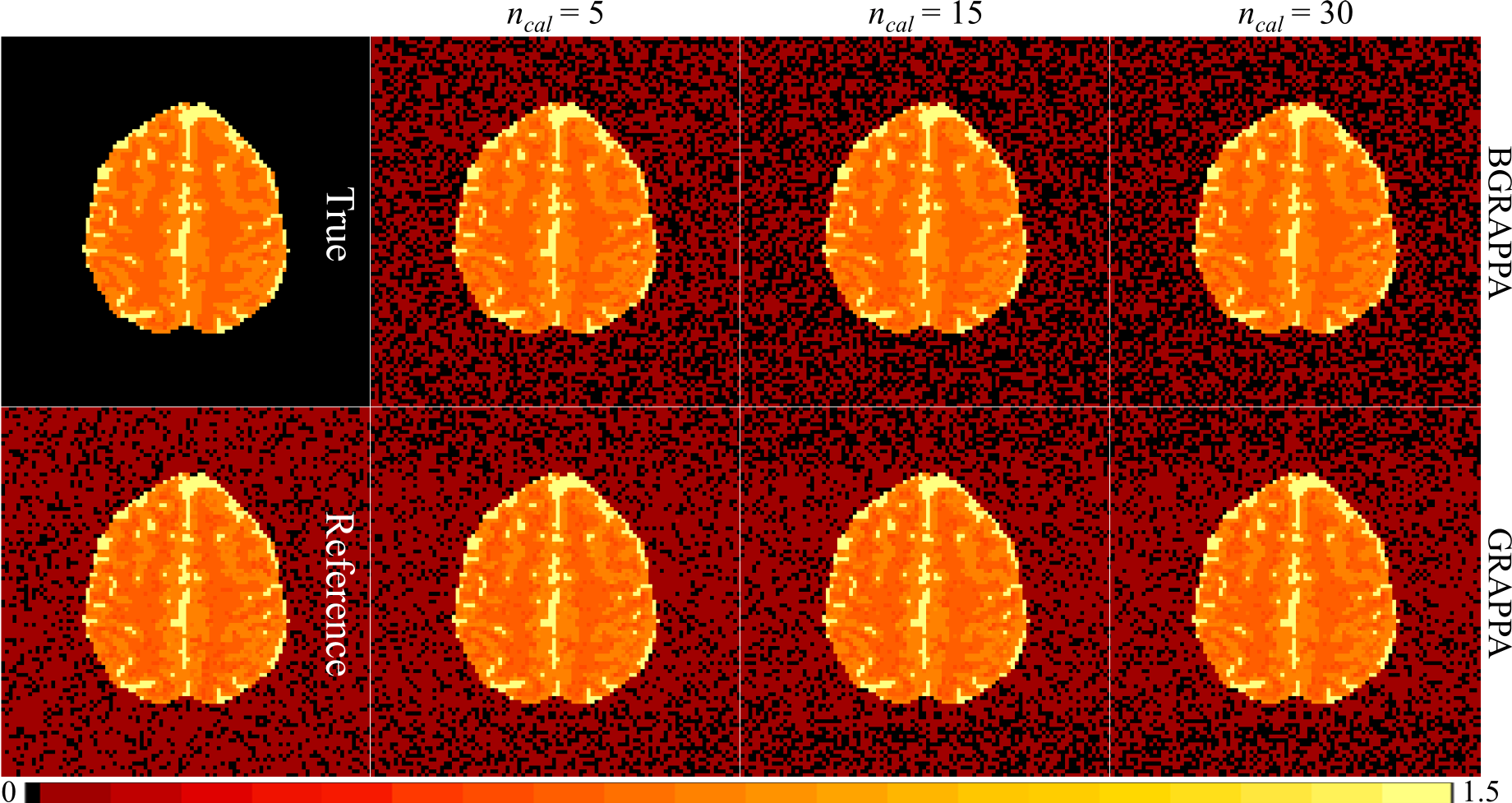}
	\caption{Reconstructed magnitude images for different number of calibration images using BGRAPPA MAP estimate (top row of the right three columns) and GRAPPA (second row of the right three columns) with the true simulated magnitude image (top left), and the reference magnitude image (bottom left).}
 \label{fig:caltestmag}
\end{figure}

\begin{figure}[!h]
	\centering
	\includegraphics[width=6in]{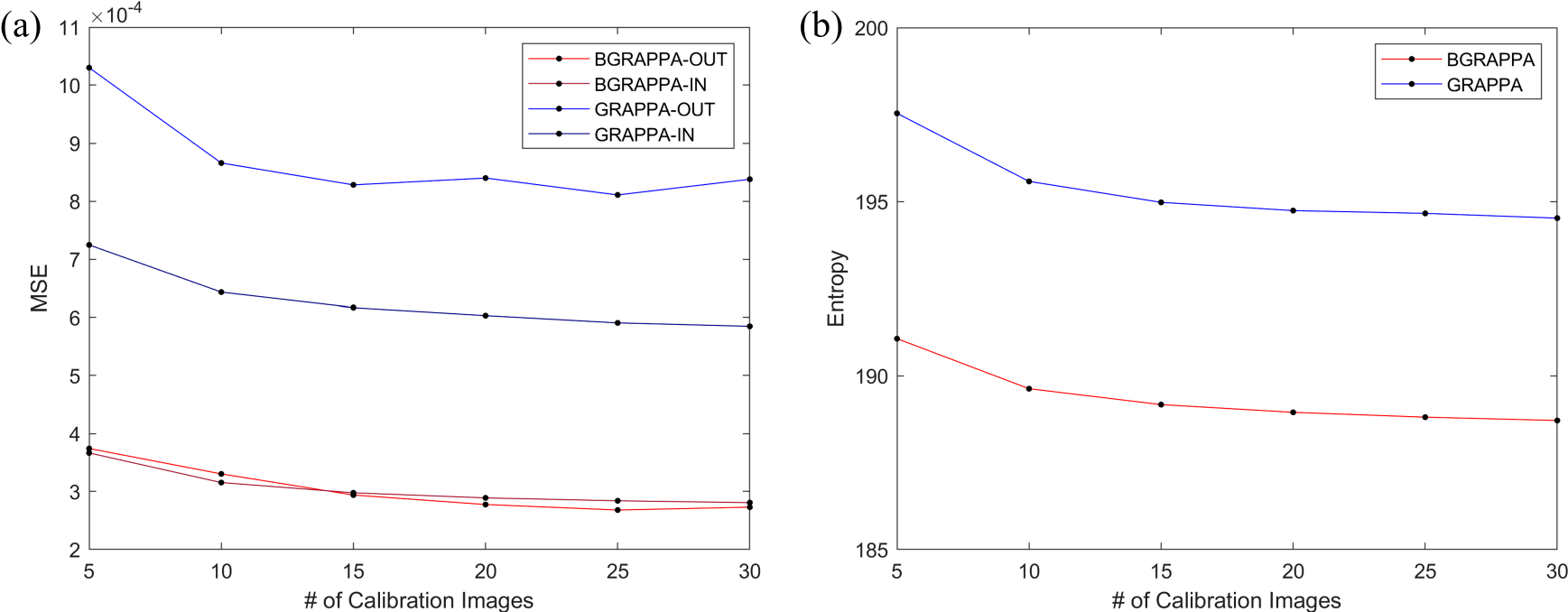}
	\caption{(a) MSE for inside and outside the brain for BGRAPPA and GRAPPA comparing both method’s reconstructed images to the true simulated magnitude image for each number of calibration images. For the MSE plot, BGRAPPA is shown in red for outside the brain (blue for GRAPPA) and dark red for inside the brain (dark blue for GRAPPA). (b) Entropy plot for BGRAPPA and GRAPPA for each number of calibration images where BGRAPPA is shown red, and GRAPPA is shown in blue.}
 \label{fig:caltestmseentropy}
\end{figure}

\begin{figure}[!h]
	\centering
	\includegraphics[width=6.4in]{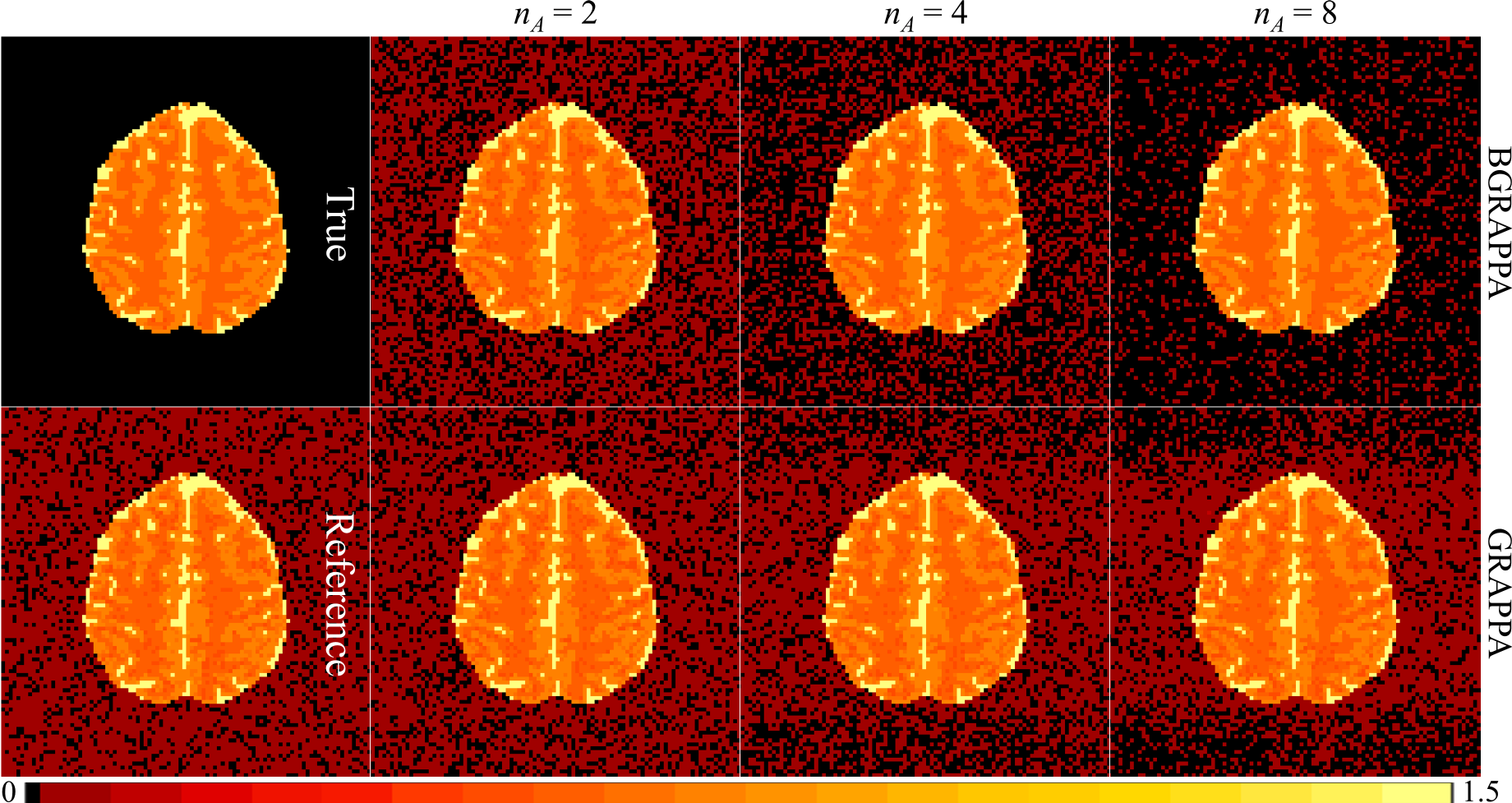}
	\caption{Reconstructed magnitude images for different acceleration factors using BGRAPPA MAP estimate (top row of the right three columns) and GRAPPA (second row of the right three columns) with the true simulated magnitude image (top left), and the reference magnitude image (bottom left).}
 \label{fig:acceltestmag}
\end{figure}

To analyze the reconstruction performance of BGRAPPA vs. GRAPPA, we first reconstructed subsampled $k$-space arrays at one time point, yielding a single unaliased image for both methods. For calibration analysis, the last ${n}_{cal}=30$ time points from the first non-task time series were utilized for hyperparameter assessment. The first time point of the 490 subsampled, simulated non-task time series with an acceleration factor of ${n}_{A}=3$, shown in Figure \ref{fig:subsamplecoilmeasurements}. The results of reconstructing the first time point in the subsampled time series using both BGRAPPA and GRAPPA are shown in Figures \ref{fig:RecSingleTP}, \ref{fig:caltestmag}, \ref{fig:caltestmseentropy}, and \ref{fig:acceltestmag}.

The prior means from the calibration information for the unacquired spatial frequency arrays ${f}_{k0}$ and the localized weights ${D}_{0}$ were used as initial values for ${f}_{k}$ and $D$. These initial values were used to generate a $\tau^{2}$ value from the posterior conditional mode from Eq. \ref{eqn:tauICM}, initializing the ICM optimization algorithm. The simulated subsampled coil $k$-space arrays were reconstructed into a single, full brain image using the BGRAPPA MAP estimate from the ICM algorithm, and traditional GRAPPA estimate. For the ICM algorithm, only three iterations were needed for estimating the parameters with a computation time of approximately half a second per time point. Figure \ref{fig:RecSingleTP} displays the true noiseless simulated image (first column) along with the reference magnitude and phase images (second column). These reference images were determined by simply averaging the full, noisy coil $k$-space arrays in the time series yielding a single spatial frequency array and then applying the IFT resulting in a full brain image. This provides us what image reconstruction would look like without applying an acceleration factor. This process is used for all of the reference images provided in the figures displaying simulated results for the remainder of the paper. Figure \ref{fig:RecSingleTP} also shows the BGRAPPA MAP unaliased image (third column), and the GRAPPA unaliased image (fourth column) for the first time point in the simulated non-task series. We can see that the joint MAP estimate from BGRAPPA and the GRAPPA estimate both produce magnitude and phase images that closely resemble the true non-aliased image in Figure \ref{fig:RecSingleTP} (left column). Visually the BGRAPPA image is slightly more accurate and less noisy than the GRAPPA image.

To quantify the differences between the true and reconstructed magnitude and phase images, we use the mean squared error, $MSE = \frac{1}{K} \sum_{j=1}^{K} \left({v}_{j}-{\overline{v}}_{j} \right)^{2}$, where $K$ is the number of voxels (either inside or outside the brain) in the full reconstructed image, ${v}_{j}$ is the reconstructed magnitude or phase value of the $j$th voxel, and ${\overline{v}}_{j}$ is the true magnitude or phase value of the $j$th voxel. This measure will indicate the accuracy of a single reconstructed image compared to the true simulated image with lower MSE indicating a more accurate reconstructed image. The MSE for BGRAPPA for inside and outside the brain was lower for the magnitude and phase images compared to GRAPPA. The MSE for the magnitude reconstructed image of GRAPPA was 114\% and 51\% higher for inside and outside the brain, respectively, compared to BGRAPPA. For the MSE of the phase reconstructed images, GRAPPA was 12\% and 3\% higher for inside and outside the brain compared to BGRAPPA.

Next, we evaluated how the number of calibration time points, ${n}_{cal}$, affected the reconstructed images. For the pre-scan calibration analysis, we fixed the acceleration factor to be ${n}_{A}=3$ for the subsampled $k$-space coil arrays of the simulated non-task time series with ${n}_{IMG}=490$ time points. Then we set the number of calibration time points to be ${n}_{cal} = 5, 10, 15, 20, 25, 30$ for separate hyperparameter assessments. After assessing the hyperparameters using each number of calibration time points, the simulated non-task time series with the subsampled coil spatial frequency arrays were reconstructed using BGRAPPA MAP and GRAPPA.

The results, displayed in Figure \ref{fig:caltestmag}, indicate that increasing the number of calibration time points does not noticeably affect the noise level inside or outside the brain for either BGRAPPA or GRAPPA. This means we can have short calibration scans and do not need to take up valubale scanner time. For each of the number of calibration time points, GRAPPA, visually, is slightly noisier than BGRAPPA. To further analyze the differences between the BGRAPPA and GRAPPA reconstructed magnitude images, the MSE and entropy for BGRAPPA and GRAPPA for each number of calibration time points were calculated to quantify this result. Entropy analyzes uncertainty and smoothness across a single image with lower entropy meaning less uncertainty throughout the image. The equation for entropy is given by $E = -\sum_{j=1}^{N} \left[\frac{{v}_{j}}{{v}_{max}}ln\left(\frac{{v}_{j}}{{v}_{max}}\right) \right]$, where $ln$ is the natural log, $N$ is the number of voxels in the full reconstructed image, ${v}_{j}$ is the reconstructed magnitude value of the $j$th voxel, and ${v}_{max}$ is the voxel intensity if all the image intensities were in one pixel \citep{atkinson1997entropy} given by ${v}_{max}=\sqrt{\sum_{j=1}^{N}{{v}_{j}}^{2}}$.

Shown in Figure \ref{fig:caltestmseentropy}a, the MSE for inside and outside the brain for the BGRAPPA MAP reconstructed magnitude images was markedly smaller than the GRAPPA reconstructed magnitude images for both inside and outside the brain. BGRAPPA also had noticeably smaller entropy values compared to the GRAPPA reconstructed magnitude images, displayed in Figure \ref{fig:caltestmseentropy}b. Lower MSE for BGRAPPA indicates a more precise reconstructed image while smaller entropy means less uncertainty with image reconstruction. For both BGRAPPA and GRAPPA, increasing the number of calibration images does not meaningfully affect the temporal variance, resulting in similar SNR for each ${n}_{cal}$ which again means we can have a short calibration scan. In all cases, the temporal variance for BGRAPPA is substantially lower than for GRAPPA, demonstrating that BGRAPPA mitigates noise in the reconstructed image. The phase of the reconstructed images for the different calibration time points can be found in Section 1 of the Supplementary Material.

Along with analysis of the number of calibration time points, we evaluated how different acceleration factors, ${n}_{A}$, affected the reconstructed magnitude and phase images. Here, we fixed the number of calibration time points to be ${n}_{cal}=30$ for hyperparameter assessment and set the acceleration factors of the non-task time series to be ${n}_{A}=2, 3, 4, 6, 8, 12$. We only show results for ${n}_{A}=2, 4, 8$ just to see how increasing the acceleration factor effects the reconstruction results. These subsampled coil $k$-space arrays with separate acceleration factors were reconstructed into full images using the BGRAPPA MAP estimate and GRAPPA, again comparing the results for both methods.

\begin{figure}[!b]
	\centering
	\includegraphics[width=6in]{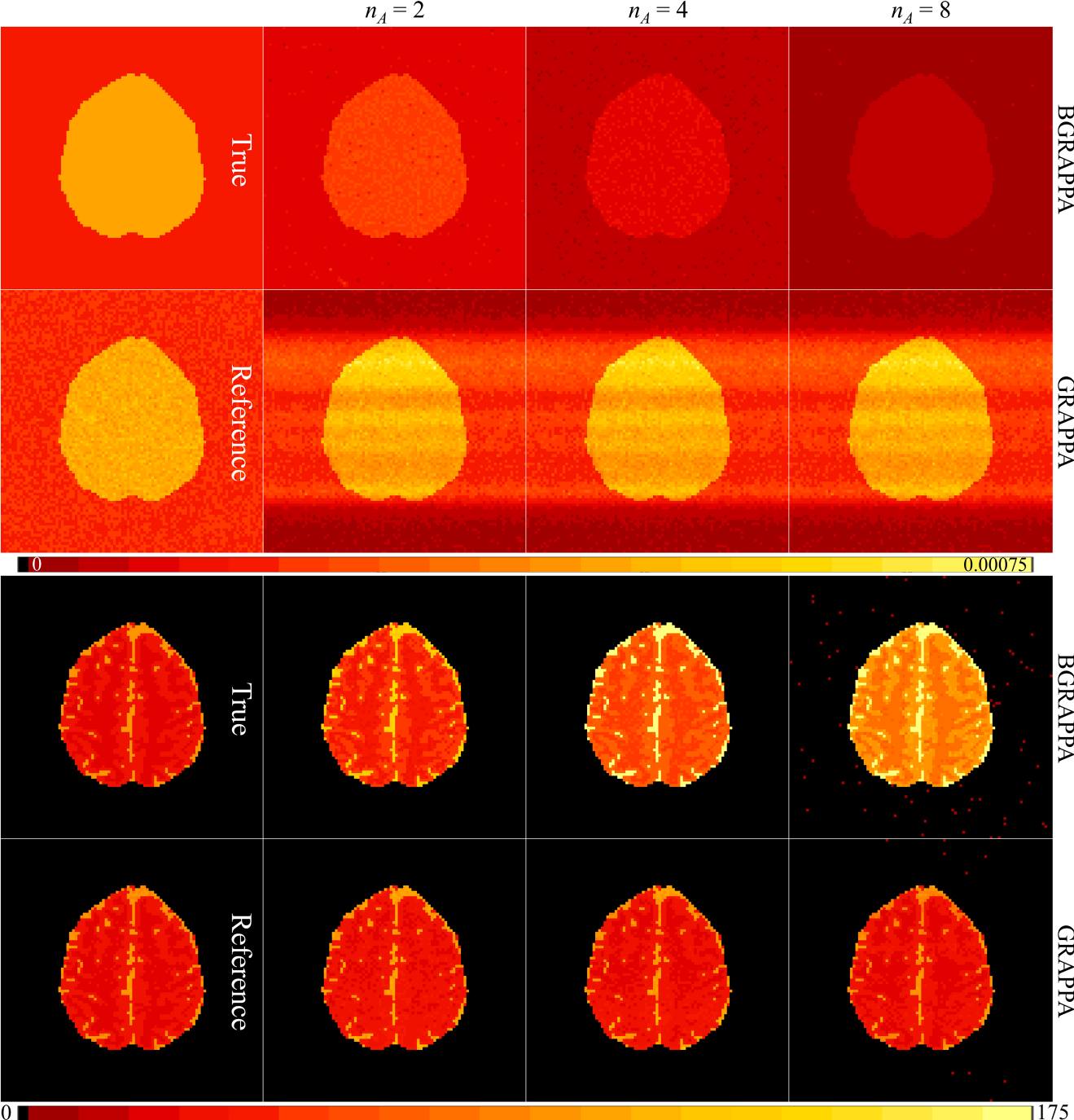}
	\caption{Temporal variance and SNR images for different acceleration factors using BGRAPPA MAP estimate (first row and third row, respectively, of the right three columns) and GRAPPA (second row and fourth row, respectively,  of the right three columns). The first column shows the true variance and SNR (first and third row, respectively) and the reference variance and SNR (second and fourth row, respectively).}
 \label{fig:acceltestvarsnr}
\end{figure}

The results, exhibited in Figure \ref{fig:acceltestmag}, showed that inside the brain of the reconstructed magnitude images from BGRAPPA and GRAPPA are negligibly affected by increasing the acceleration factor with BGRAPPA visually slightly more accurate. The noise level outside the brain for BGRAPPA does decrease as the acceleration factor increases while it only slightly decreases for GRAPPA. Again, the GRAPPA magnitude reconstructed images have slightly more noise than the BGRAPPA magnitude reconstructed images. The phase of the reconstructed images for the different acceleration factors can be found in Section 1 of the Supplementary Material.

In Figure \ref{fig:acceltestvarsnr}, we examine the temporal variance of the reconstructed time series and the SNR images for BGRAPPA, GRAPPA, the true images, and the reference images when no acceleration factor is applied. The reference reconstruction is simply averaging the full coil $k$-space arrays and applying the IFT to get full brain images for the full time series. The temporal variance for BGRAPPA decreases and for GRAPPA increases as the acceleration factor increases (first and second row of Figure \ref{fig:acceltestvarsnr} of the right three columns). The temporal variance, overall, from Figure \ref{fig:acceltestvarsnr} for BGRAPPA is substantially lower than GRAPPA, the theoretically true variance (first row, first column), and the reference (second row, first column), showing that BGRAPPA reduces the noise through of the reconstructed time series. This also leads to higher SNR for BGRAPPA compared to GRAPPA (third and fourth row of Figure \ref{fig:acceltestvarsnr} of the right three columns), the true SNR (first column, third row), and the reference SNR (first column, fourth row). The average of the BGRAPPA reconstructed time series was also taken and the result magnitude image looks similar to the true simulated magnitude image.

\subsection{Task Activation Model}\label{subsec:SimDataTaskAct}
In task-based fMRI, the non-task reconstructed images create a baseline value for each voxel. This yields an intercept only simple linear regression model $y={\beta}_{0}+\varepsilon$ where $y$ is the magnitude of the reconstructed voxel value. By adding in task activation to select images in the series of images, we have a simple linear regression model $y={\beta}_{0}+x{\beta}_{1}+\varepsilon$ for the unaliased voxel values. In this regression, ${\beta}_{0}$ is the baseline voxel value from the non-task reconstructed images determining the SNR $={\beta}_{0}/\sigma$, as demonstrated in the previous subsection. The ${\beta}_{1}$ value is the estimated task related signal increase from ${\beta}_{0}$ determining the contrast-to-noise ratio CNR $={\beta}_{1}/\sigma$. The vector $x\in{\{0,1\}}^{{n}_{IMG}}$, where ${n}_{IMG}$ is the number of reconstructed images in the series, is a vector such that the zeros correspond to the images in the series without task activation and the ones correspond to the images with task activation. We can write this regression as $y=XB+\varepsilon$, where $X=\left[1,\hspace{2mm}x\right]\in\mathbb{R}^{{n}_{IMG}\times2}$ and $B=[{\beta}_{0},{\beta}_{1}]'$.

The task is not usually visible on the reconstructed images since the CNR is typically much lower than the SNR. Instead, a right-tailed $t$-test is carried out with ${\beta}_{1}\leq0$ as the null hypothesis and ${\beta}_{1}>0$ as the alternative. The reason for the one-sided hypothesis test is because we only anticipate an increased signal from the task activation. To simulate added task, a ${\beta}_{1}=0.045$ magnitude-only signal increase is added to select voxels of the true noiseless non-task image. This added task activation is located in the left motor cortex to resemble the region of interest (ROI) of brain activity from the fMRI unilateral right-hand finger tapping experiment used in the Section \ref{sec:ExpData} \citep*{karaman2014fmri}.

Similar to magnitude-only task activation, we can also use the phase images for task detection. A simulated phase task of $\pi/120$ was also added to the simulated true simulated task image. A simple linear regression model, $\phi={\theta}_{0}+{\theta}_{1}x+\epsilon$, can be used for the phase activation as well. In this regression, $\phi$ is the phase of the reconstructed voxel, ${\theta}_{0}$ is the baseline phase voxel value from the non-task reconstructed images, and ${\theta}_{1}$ is the estimated increase from ${\theta}_{0}$. We then use a one-tailed $t$-test, $t={\hat{\theta}}_{1}/SE({\hat{\theta}}_{1})$, to determine which voxels contain statistically significant ${\theta}_{1}$ values indicating which voxels experience phase task activation \citep{rowe2007phaseonly}.

\subsection{FMRI Spatial Frequency Data}\label{subsec:SimDatafMRIGen}
A noiseless task image was used along with the noiseless non-task image to create a series of ${n}_{TR}=510$ simulated full coil images for one slice mimicking real-world fMRI data. The simulated task activation was designed to mimic tapping of the subject’s right fingers leading to activity in the left motor cortex which becomes our ROI for analyzing task detection in this experiment, as mentioned above. Knowing this, artificial signal increase was added to the voxels in the ROI (as mentioned in Subsection \ref{subsec:SimDataTaskAct}) for task images.

The true images were multiplied by the same ${n}_{C}=8$ coil sensitivity maps used for the non-task simulated time series (as illustrated in Figure \ref{fig:weightedcoils}), and then the series of images were Fourier transformed in full coil $k$-space arrays. This series was also generated by adding separate $N(0,0.0036{n}_{y}{n}_{x})$ noise to the real and imaginary parts of the full coil $k$-space arrays and were then inverse Fourier transformed back into full coil images, yielding a CNR of 0.75. To simulate our real-world fMRI experimental process, the first 20 time points in the series 20 were non-task. The scaling for the first few images in the fMRI simulated data is similar to that outlined in Subsection \ref{subsec:SimDataGen} for each of the tissue types. The initial 20 non-task time points are followed by 16 epochs alternating between 15 non-task and 15 task time points. An epoch is a stimulation period with time points of the subject at rest (non-task) and the subject performing an action or task. The series culminated with 10 non-task time points producing the simulated fMRI series of ${n}_{TR}=510$ images. To mimic the fMRI experiment in Section \ref{sec:ExpData}, the first 20 time points were omitted leaving 490 time points in the series. This series is transformed into the spatial frequency domain and then subsampled according to the acceleration factor to simulate subsampling of a real fMRI experiment. The last ${n}_{cal}$ full coil FOV time points in the second non-task time series from Subsection \ref{subsec:SimDataGen} were utilized as full FOV coil calibration $k$-space arrays to assess the hyperparameters. For this simulation, we evaluate both BGRAPPA and GRAPPA using ${n}_{cal}=5, 10, 15, 20, 25, 30$ calibration time points. The results for the different calibration time points were similar to that of Subsection \ref{subsec:SimDataResults} where the different number of calibration time points had negligible affects on the results. Different acceleration factors of ${n}_{A}=2, 3, 4$ were also tested in this simulated fMRI experiment and are shown in the next subsection.

\subsection{FMRI Reconstruction Results}\label{subsec:SimDatafMRIResults}

\begin{figure}[!h]
	\centering
	\includegraphics[width=6.2in]{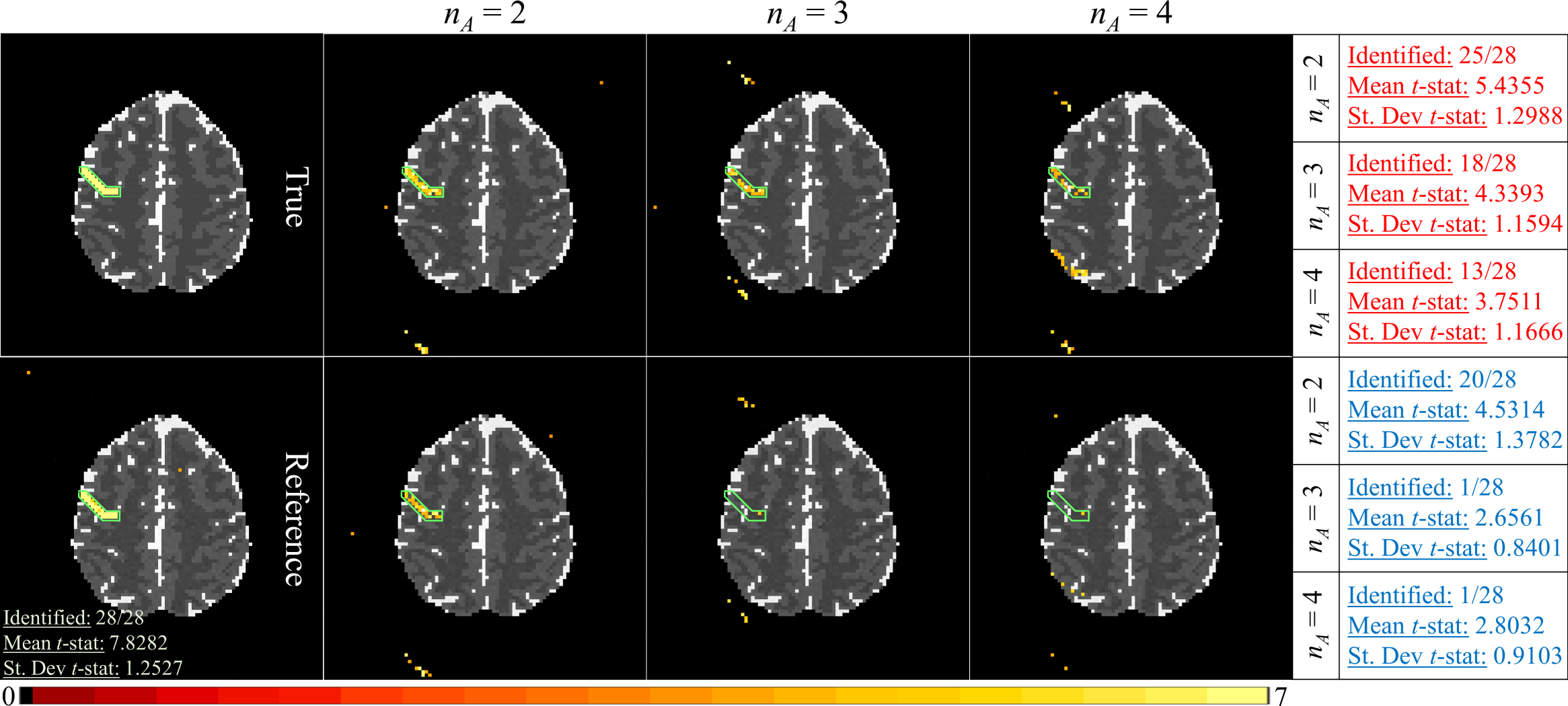}
	\caption{Statistically significant voxels in the ROI using FDR for BGRAPPA reconstructed images (first row of the right three columns), significant voxels in the ROI using FDR for GRAPPA (second row of the right three columns), and analysis of the $t$-statistics in the boxes on the right with BGRAPPA in red and GRAPPA in blue. The true (first row) and the reference (second row) magnitude-only task activation is shown in the first column with the analysis of $t$-statistics of the reference reconstruction shown in the image.}
 \label{fig:RecTaskActAccelTest}
\end{figure}

The hypothesis test described in Subsection \ref{subsec:SimDataTaskAct} was utilized to determine voxels with a statistically significant magnitude-only signal increase. The statistically significant voxels for different acceleration factors were analyzed for the BGRAPPA MAP reconstructed time series and the GRAPPA reconstructed time series using the 5\% false discovery rate (FDR) threshold procedure (\citealp{benjamini1995fdr}; \citealp*{genovese2002thresh}; \citealp{logan2004thresh}). The ROI here consists of 28 voxels located in the left motor cortex.

\begin{figure}[!b]
	\centering
	\includegraphics[width=6.2in]{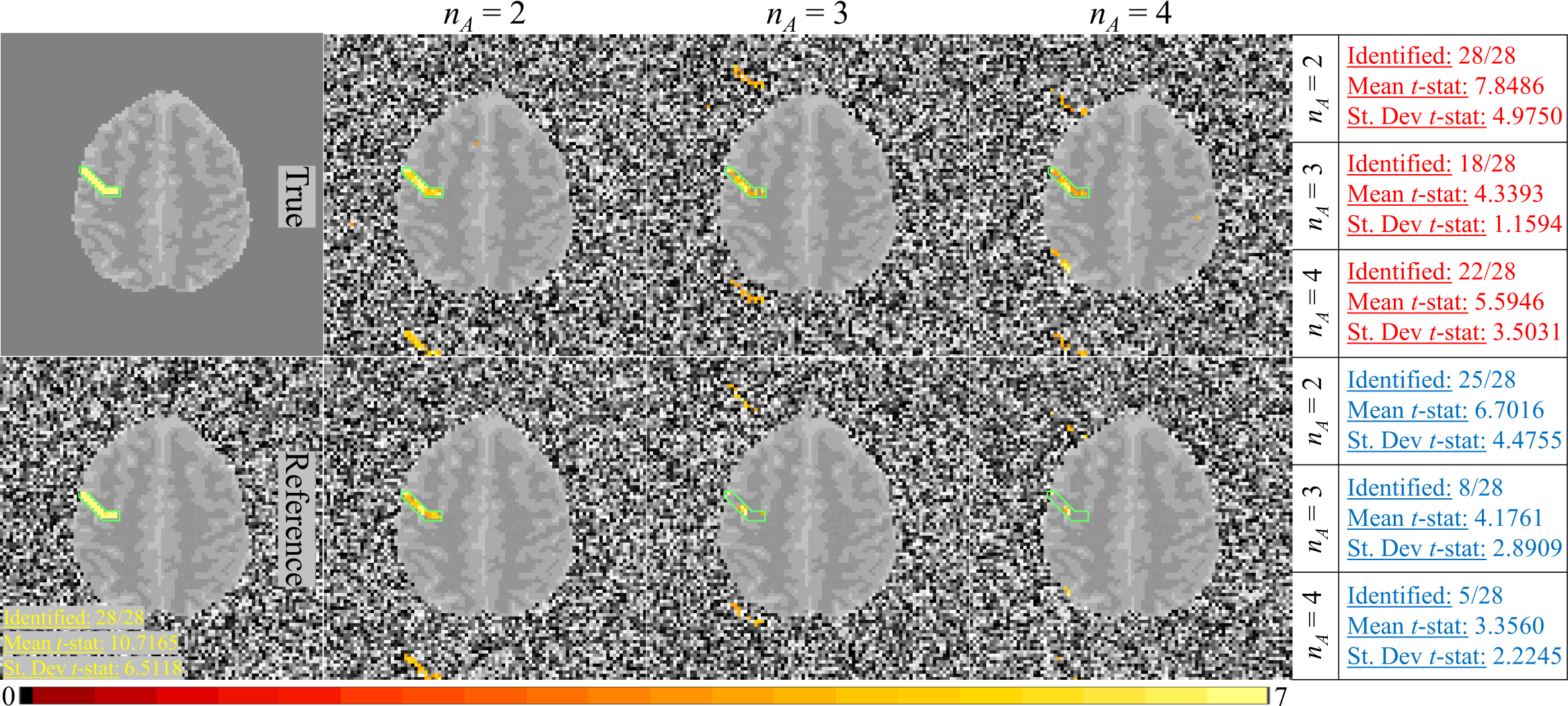}
	\caption{Statistically significant voxels in the ROI using FDR for BGRAPPA phase reconstructed images (first row of the right three columns), significant voxels in the ROI using FDR for GRAPPA (second row of the right three columns), and analysis of the $t$-statistics in the boxes on the right with BGRAPPA in red and GRAPPA in blue. The true (first row) and the reference (second row) phase-only task activation is shown in the first column with the analysis of $t$-statistics of the reference reconstruction shown below in the image.}
 \label{fig:RecTaskActAccelTestphase}
\end{figure}

Figure \ref{fig:RecTaskActAccelTest} shows the statistically significant magnitude-only voxels from the BGRAPPA MAP reconstructed time series (first row of the right three columns) and the GRAPPA reconstructed time series (second row of the right three columns) for the different acceleration factors compared true and reference activations in the first column. Figure \ref{fig:RecTaskActAccelTest} also summarizes the $t$-statistics in the ROI for each acceleration factor. BGRAPPA identified more statistically significant voxels in the ROI for each acceleration factor. For acceleration factors of 3 and 4, the task activation is virtually undetected using the GRAPPA method. The mean value for the $t$-statistics was also substantially higher for BGRAPPA compared to GRAPPA, demonstrating that BGRAPPA has a stronger task detection power. Increasing the acceleration factor decreases number of voxels identified and the mean of the $t$-statistics for both BGRAPPA and GRAPPA, but much more activation is captured from BGRAPPA than GRAPPA.

Using the 5\% FDR threshold, Figure \ref{fig:RecTaskActAccelTestphase} shows phase activation for BGRAPPA and GRAPPA reconstructed time series using acceleration factors of 2, 3, and 4. Like the BGRAPPA reconstructed magnitude images, we can see that it captures the simulated task activation in the ROI for each acceleration factor. For GRAPPA, the phase task activation is captured using an acceleration of 2, but noticeably diminishes when applying acceleration factors of 3 and 4. With higher mean $t$-statistic values for BGRAPPA, this indicates that BGRAPPA has more power in phase task detection. Phase activation is a topic of study as previously described in Subsection \ref{subsec:SimDataResults}.

\section{Experimental Data}\label{sec:ExpData}
\subsection{Data Description}\label{subsec:ExpDataHumanSubject}
A 3.0 T General Electric Signa LX magnetic resonance imager was used to conduct an fMRI experiment on a single subject. A right-handed finger-tapping task was performed in a block design with an initial 20 s rest followed by 16 epochs with 15 s off (rest state) and 15 s on (task performed). The experiment was concluded with 10 s of rest giving us a series of ${n}_{TR}=510$ repetitions with each repetition being 1 s, a flip angle of 90° and an acquisition bandwidth of 125 kHz. The data set consists of nine 2.5 mm thick axial slices with ${n}_{C}=8$ receiver coils that have a 96$\times$96 dimension for a 24 cm full FOV, with a posterior to anterior phase encoding direction. For this paper, the time series for all nine slices was used to analyze the effects of applying acceleration factors of ${n}_{A}=2, 3, 4$ for both BGRAPPA and GRAPPA, but only the time series of the second slice is shown. Note that the simulation study in Section \ref{sec:SimData} directly mimic this experimental data.

Typically, the magnetic fields in an fMRI experiment will induce a drift in the phase over time which we correct before reconstruction to give us a stable phase through time. Once the phase was corrected, the last ${n}_{cal}=30$ full $k$-space arrays of a non-task series of ${n}_{TR}=510$ time points performed on the subject were used for hyperparameter assessment. The fMRI experimental series described above was used for fMRI analysis. The first 20 images of each series were discarded due to varying echo times and magnetization stability. Like the simulation study, the subsampled coil $k$-space arrays came from artificially skipping lines in the full coil $k$-space arrays of the fMRI experimental time series, mimicking the effect of actually subsampling the coil $k$-space arrays. Before subsampling the time series, a reference image (left image in Figure \ref{fig:ExpRecAccelTest}) was produced by averaging the ${n}_{C}=8$ full coil spatial frequency arrays at the first time point. This provides a magnitude and phase image with which to compare to GRAPPA and our proposed BGRAPPA.

\subsection{Experimental Results}\label{subsec:ExpDataResults}
\begin{figure}[!h]
	\centering
	\includegraphics[width=6.4in]{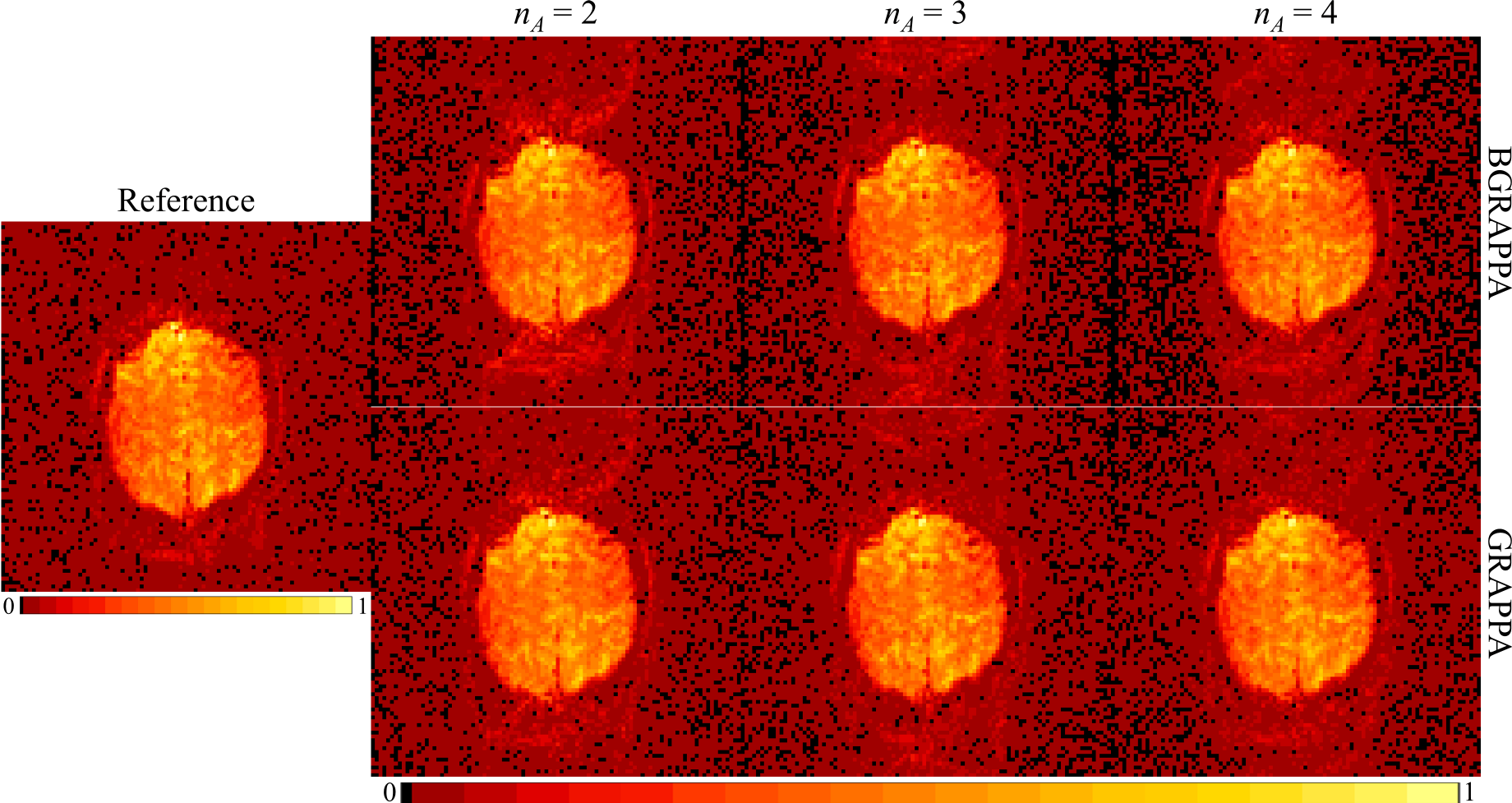}
	\caption{BGRAPPA MAP unaliased non-task magnitude images for each acceleration factor (first row  of the right three columns) using the ICM algorithm, and GRAPPA unaliased non-task magnitude images for each acceleration factor (second row  of the right three columns) with the magnitude reference image (left column).}
 \label{fig:ExpRecAccelTest}
\end{figure}

Similar to the process for the simulated data described in Section \ref{sec:SimData}, the unacquired spatial frequencies at each time point in the entire time series of subsampled coil $k$-space arrays were estimated using BGRAPPA and GRAPPA separately. The estimated full coil $k$-space arrays were then averaged together and transformed into image space resulting in a single composite brain image. Figure \ref{fig:ExpRecAccelTest} displays the BGRAPPA MAP reconstructed images (top row) and the GRAPPA reconstructed images (bottom row) of the first time point of the 490 images using acceleration factors 2, 3, and 4. Just as the simulated results in Figure \ref{fig:acceltestmag} demonstrated, the BGRAPPA reconstruction method in Figure \ref{fig:ExpRecAccelTest} produced visually similar magnitude reconstructed images compared to GRAPPA reconstruction but with slightly less noise.

MSE was again utilized to quantify the differences between the reference image and reconstructed images. The MSE for inside the brain for GRAPPA was approximately 12\%, 10\%, and 3\% higher for each acceleration factor, respectively, compared to BGRAPPA. GRAPPA having a larger MSE inside the brain for each acceleration factor, respectively, reflects decreased noise from BGRAPPA versus GRAPPA. The entropy for BGRAPPA (214.1026, 207.5331, and 204.1746, respectively) was also lower than the entropy for GRAPPA (216.0362, 212.3556, and 210.3667, respectively) indicating that the BGRAPPA reconstructed images are more smooth. The phase of the experimental reconstructed images for the different acceleration factors can be found in Section 2 of the Supplementary Material.

\begin{figure}[!b]
	\centering
	\includegraphics[width=6.4in]{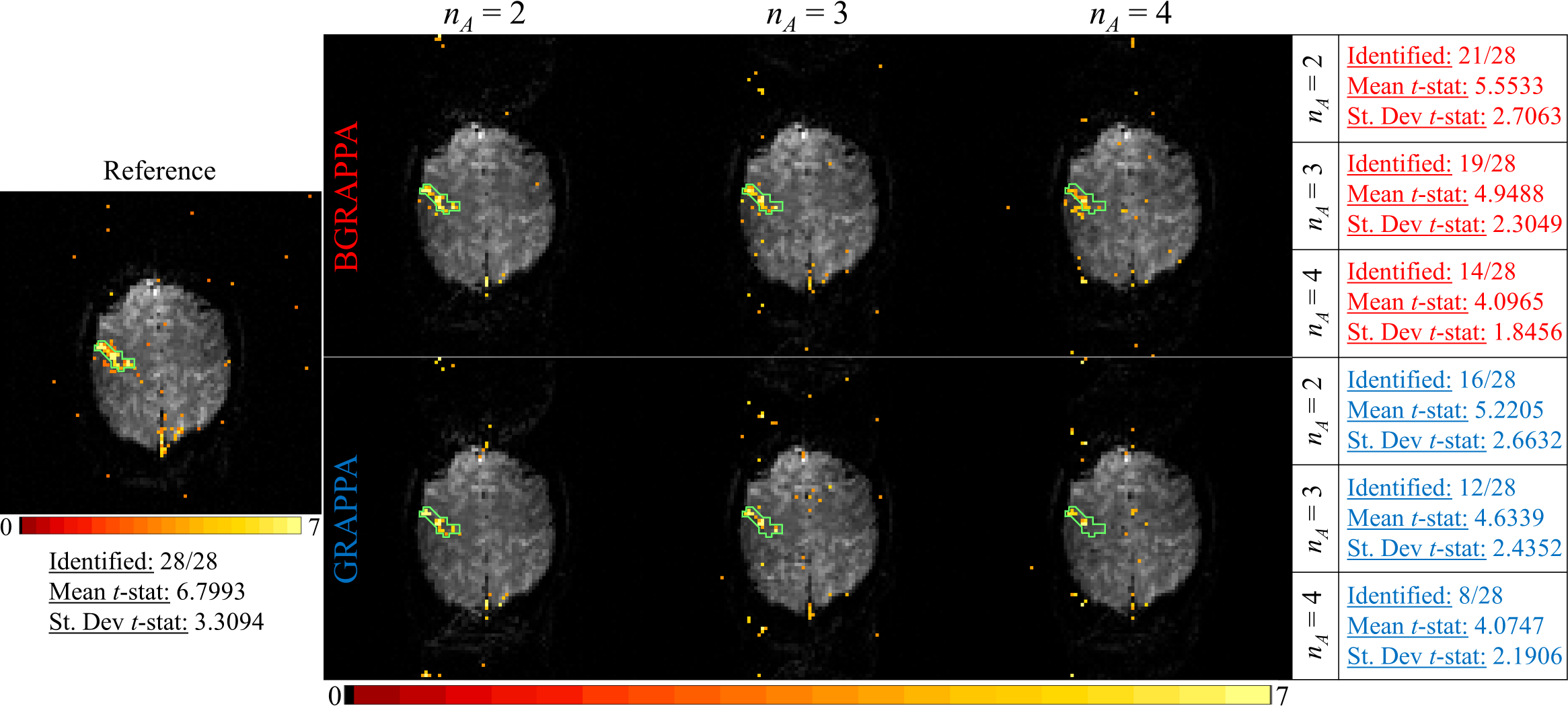}
	\caption{Statistically significant voxels in the ROI using FDR for BGRAPPA reconstructed images (first row of the right three columns) for three different acceleration factors, significant voxels in the ROI using FDR for GRAPPA (second row of the right three columns) for three different acceleration factors, and analysis of the $t$-statistics to the right of the images with BGRAPPA in red and GRAPPA in blue. The reference magnitude-only task activation is shown on the left column with the analysis of $t$-statistics below in black.}
 \label{fig:ExpRecTaskActAccelTest}
\end{figure}

For the detection of magnitude task activation, the hypothesis test outlined in Subsection \ref{subsec:SimDataTaskAct} was carried out. Figure \ref{fig:ExpRecTaskActAccelTest} shows the statistically significant voxels under BGRAPPA (top row) and GRAPPA (bottom row) reconstruction. The images for the statistically significant voxels in Figure \ref{fig:ExpRecTaskActAccelTest} for both methods use the 5\% FDR threshold. Voxels outside the brain are usually masked out meaning the statistically significant voxels shown outside the brain in Figure \ref{fig:ExpRecTaskActAccelTest} would typically be discarded. Figure \ref{fig:ExpRecTaskActAccelTest} also summarizes the $t$-statistics with BGRAPPA (red) and GRAPPA (blue). BGRAPPA correctly detected more voxels than GRAPPA as task activation in the ROI for all three acceleration factors. Our BGRAPPA approach also had a much higher mean $t$-statistic and a lower standard deviation for all the acceleration factors.

\section{Discussion}\label{sec:Discussion}
Parallel imaging techniques such as GRAPPA \citep{griswold2002grappa} have utilized subsampling of $k$-space to reduced the acquisition time for MR imaging. This allows practitioners to reconstruct higher resolution images, decrease the time between each image, increase the number of images or slices in an fMRI experiment, or a combination of both in the same time as fully sampled $k$-space, depending on the acceleration factor. The acceleration factor in an fMRI experiment is determined by how important time is in completing a scan. The number of coils used in an experiment is dependent on the coil configurations that facility possesses.

Applying an acceleration factor in an fMRI experiment can significantly reduce the acquisition time of spatial frequency arrays and volume images, but taking the IFT of the subsampled $k$-space yields aliased images. GRAPPA parallel image reconstruction estimates the unacquired spatial frequencies that are skipped during the acquistion of the subsampled $k$-space arrays yielding full FOV coil spatial frequency arrays. However, GRAPPA has its drawbacks which include low image quality, low SNR, and weakened task detection power at higher acceleration factors. Hence we introduce a Bayesian approach to estimate the unacquired spatial frequencies. Using more available information from the calibration spatial frequencies to assess the hyperparameters, our proposed approach successfully reconstructed a series of simulated non-task images without any aliasing artifacts. The BGRAPPA reconstructed images were shown to more accurately reconstruct the truth compared to GRAPPA. The number of calibration time points had minimal effect on the GRAPPA reconstructed images and its performance against BGRAPPA reconstructed images. The results also indicated that the different acceleration factors had little effect on the reconstruction of the images but did have lesser task detection power for both methods. Our BGRAPPA approach had better performance when detecting the signal increase in the voxels that experienced task activation which is demonstrated from both the simulated and experimental data.

For this paper, only the MAP estimate using the ICM algorithm was used to reconstruct the time series both the simulated and experimental data. Since we have posterior conditionals for each of the parameters, this allows us to use other estimation techniques such as the MCMC Gibbs sampling method. We chose not to present this method due to the Gibbs sampler being more computationally expensive when running a long series of images so it is not be as practical to use compared to evaluating the MAP estimate. This does not mean there is no value in running a Gibbs sampler, as it has the additional benefit of quantifying uncertainty. For instance, it can be utilized on a shorter series of images, provide us more statistical information about any voxel, hypothesis testing between two reconstructed images, or identifying which voxels are outside the brain for masking. We could also hybridize the ICM and Gibbs sampler where we start with a few iterations of the ICM algorithm followed by a short, no-burn Gibbs sampler. Our Bayesian approach allows for more options of how to run an fMRI experiment based on the objective of the scan compared to GRAPPA.

In this paper, a magnitude-only and phase-only activation model was utilized to detect task activation. Due to the high noise in the experimental data set, there were no phase active voxels as there were in the idealized simulation. Since the reconstructed images are complex-valued, our proposed model is expected to be applicable for complex activation models for task detection \citep{rowe2004fmri,rowe2005complex,rowe2009magphasesignal} as well as magnetic field mapping. Further, our proposed procedure can also be repeated for vertical aliasing as opposed to the horizontal aliasing used here.

\section*{Supplementary Materials}\label{sec:Supplement}
The supplement to this paper provides additional results for phase reconstructed images, subsampling the calibration time points for separate hyperparameter assessment, and details about the experimental data.

\vfill\eject

\bibliographystyle{unsrtnat}
\bibliography{references}  






\end{document}


\maketitle

\section{Simulated Results}\label{SimRecSupplement}

\begin{figure}[!b]
	\centering
	\includegraphics[width=6.1in]{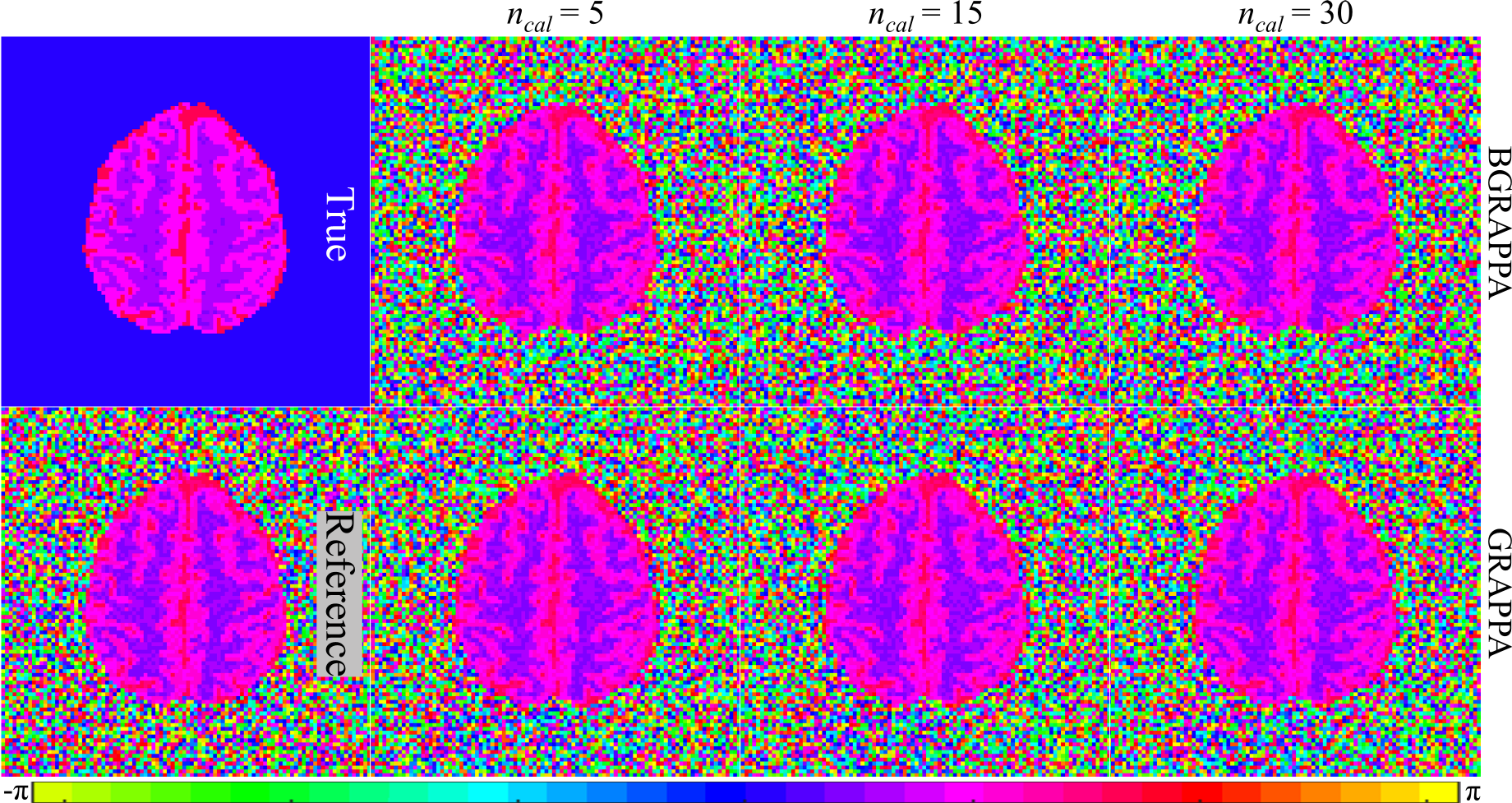}
	\caption{Reconstructed phase images for different number of calibration time points using BGRAPPA MAP estimate (top row of the right three columns) and GRAPPA (second row of the right three columns) with the true simulated phase image (top left), and the reference phase image (bottom left). Due to the circular nature of phase angles, the color bar for the phase images have wrap-around.}
 \label{fig:caltestphase}
\end{figure}

Since the reconstructed images are complex-valued, we are able to examine the phase images. Even though phase images are typically discarded in reconstruction analysis, there can be value information that you could get from the phase that may not be in the magnitude images \citep{rowe2004fmri,rowe2005complex,rowe2007phaseonly,rowe2009magphasesignal}. The phase of the reconstructed images for a different number of calibration images is shown in Figure \ref{fig:caltestphase} for BGRAPPA (top row of the right three columns) and GRAPPA (bottom row of the right three columns). The images in Figure \ref{fig:caltestphase} indicate that increasing the number of calibration images has little to no effect on the phase of the reconstructed images for both BGRAPPA and GRAPPA. Both methods produce phase images that resemble the true simulated phase with GRAPPA being visually slightly less accurate and noisier than BGRAPPA.

\begin{figure}[!h]
	\centering
	\includegraphics[width=6.1in]{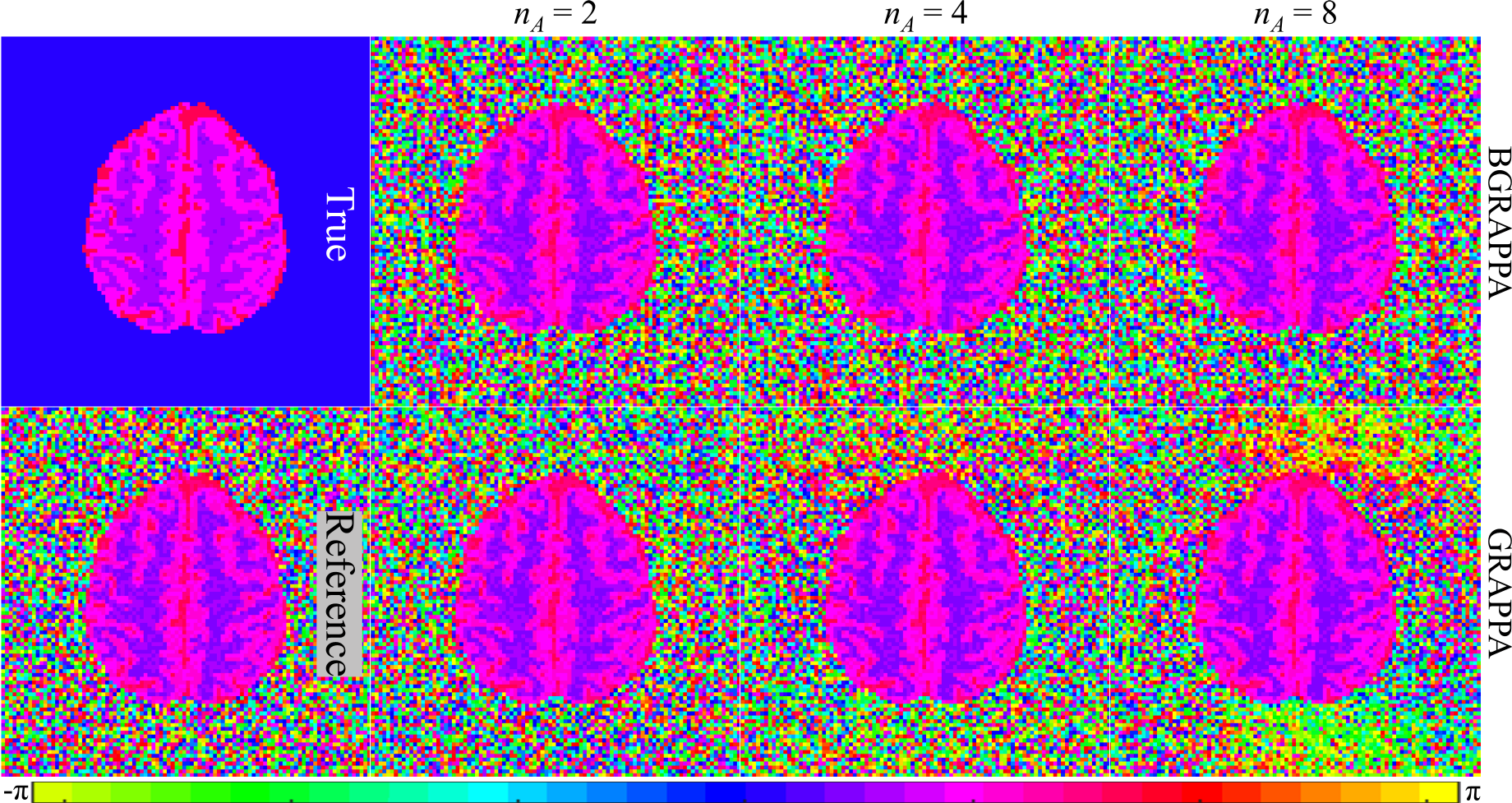}
	\caption{Reconstructed phase images for different acceleration factors using BGRAPPA MAP estimate (top row of the right three columns) and GRAPPA (second row of the right three columns) with the true simulated phase image (top left), and the reference phase image (bottom left). Due to the circular nature of phase angles, the color bar for the phase images have wrap-around.}
 \label{fig:acceltestphase}
\end{figure}

Figure \ref{fig:acceltestphase} displays the phase of reconstructed images for both BGRAPPA (top row of the right three columns) and GRAPPA (bottom row of the right three columns) using acceleration factors of 2, 4, and 8. The phase images inside the brain for BGRAPPA and GRAPPA are unaffected by the different acceleration factors. For GRAPPA, the phase images appear to have a minor aliasing effect outside the brain as the acceleration factor increases with BGRAPPA having visually lower noise and higher accuracy than GRAPPA. Note that as ${n}_{A}$ increases, the GRAPPA phase images exhibit homogeneous values outside the brain when they should be random.

For estimating priors of BGRAPPA, we use up to 30 calibration time points which are utilized to assess the hyperparameters. This means the same prior information is used at each time point when reconstructing the fMRI time series which could potentially lead to correlation with previously aliased voxels or task leakage (false detection of task in voxels that were previously aliased). To possibly mitigate this, we can randomly subsample the calibration spatial frequency arrays used at each time point in the fMRI time series. This means different hyperparameters are applied at each time point to the reconstruction of the subsampled $k$-space time series. We can also change the prior scalars ${n}_{k}$ and ${n}_{w}$ in the posterior conditional mode equations to be less than the number of calibration time points which decreases the weight of the prior information incorporated in the reconstructed images.

\begin{figure}[!h]
	\centering
	\includegraphics[width=6.2in]{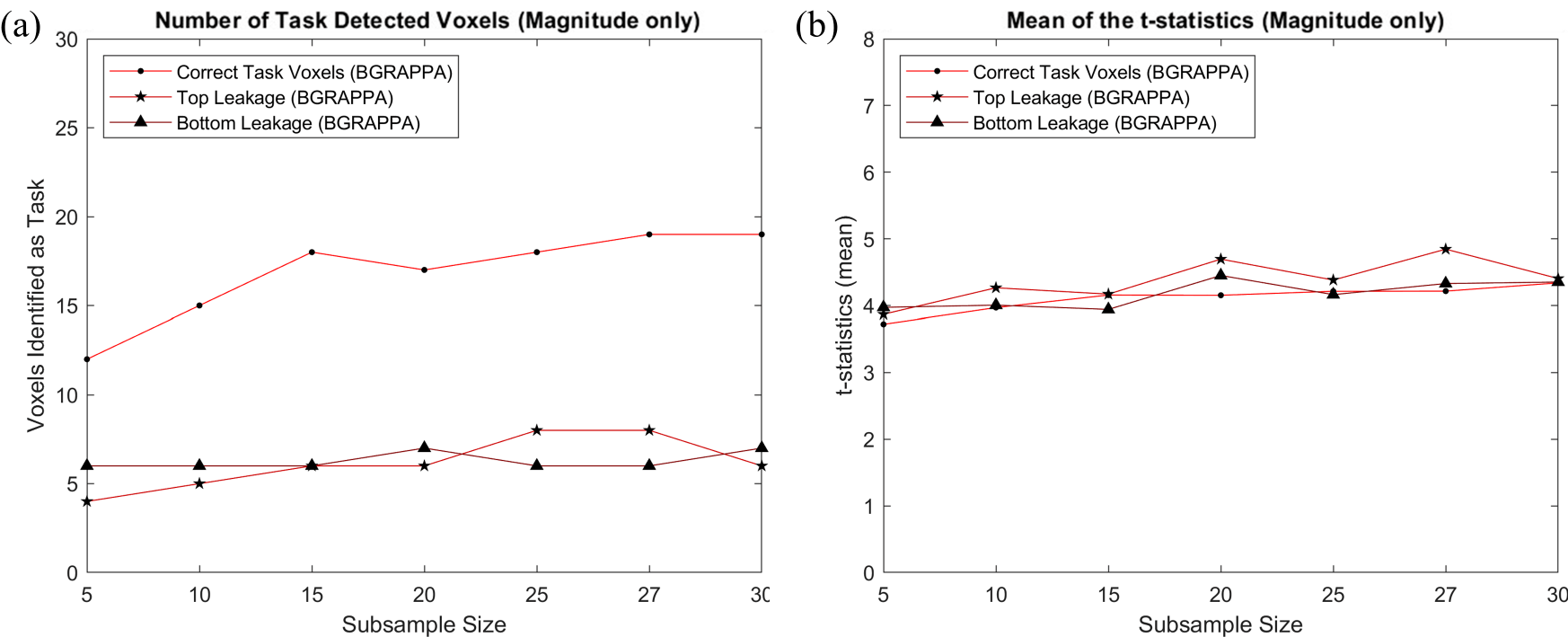}
	\caption{(a) Plot for number of voxels detected as task activation in the ROI and location of the potential task leakage (with ${n}_{A} = 3$) for each subsample size for bootstrapping the calibration image. (b) Plot for mean of the $t$-statistics of voxels in the ROI and location of the potential task leakage (with ${n}_{A} = 3$) for each subsample size for bootstrapping the calibration images. The lines with the black dots indicate the lines for analysis inside the ROI, the lines with the black stars indicate analysis for the top task leakage, and the lines with the black triangles lines for analysis of the bottom task leakage.}
 \label{fig:BGRAPPAMagActCNR075_subtest_n01_tstat_voxidentify}
\end{figure}

Here, we analyzed how the same subsampling parameters effect task detection for BGRAPPA fixing the prior scalars ${n}_{k}=1$ and ${n}_{W}=1$. Figure \ref{fig:BGRAPPAMagActCNR075_subtest_n01_tstat_voxidentify}a displays the number of voxels identified as task in the ROI along with the top and bottom task leakage from unaliasing. Figure \ref{fig:BGRAPPAMagActCNR075_subtest_n01_tstat_voxidentify}b illustrates the mean of the $t$-values in the ROI, the voxels in the top leakage, and the voxels in the bottom leakage. For both plots in Figure \ref{fig:BGRAPPAMagActCNR075_subtest_n01_tstat_voxidentify}, the lines with black dots are the values for the ROI, the lines with the black stars are the values for the top leakage region, and the lines with the black triangles are the values for the bottom leakage region.

As demonstrated in Figure \ref{fig:BGRAPPAMagActCNR075_subtest_n01_tstat_voxidentify}a, the number of magnitude task voxels correctly identified does not drastically change (except for subsampling of 5 calibration time points). Subsampling the calibration time points does not markedly decrease the number of voxels incorrectly identified as task in the leakage areas either. Corresponding results for this method are exhibited in analyzing the mean $t$-statistics shown in Figure \ref{fig:BGRAPPAMagActCNR075_subtest_n01_tstat_voxidentify}. Along with the identification of task voxels, introducing subsampling does not affect the mean of the $t$-values for the ROI or the region of task leakage compared to the normal BGRAPPA method. These results show that subsampling does not affect the power of task detection for BGRAPPA.

\section{Experimental Data and Results}\label{ExpRecSupplement}
For each volume image in the experimental series, a time dependent echo time, ${TE}_{t}$, consisted of three parts. The first part was fixed to have a value of $TE = 42.7$ ms at the first 10 time points. In the second part, the next five $TE$ values were an equally spaced interval of values 42.7 ms, 45.2 ms, 47.7 ms, 50.2 ms, and 52.7 ms and was repeated for another 5 time points. For the final part, the last 490 time points were fixed at 42.7 ms. To account for ${T}_{1}$ effects and varying echo times, the center row of $k$-space for each TR in each receiver coil was acquired with three navigator echoes which is used to correct any potential Nyquist “ghosting.” The additional rows of $k$-space were integrated to estimate and adjust the error in the center frequency and group delay offsets between the odd and even lines of $k$-space \citep*{nencka2008navigator}.

Since the phase drifts over time, we must correct it to get a stable phase through time. Before correcting the phase, the full non-task and full task experimental time series were appended to be one long time series. The angular phase temporal mean of the long time series is calculated and subtracted for each voxel time-series. A local spatial second order polynomial was spatially fit to the resultant difference of the voxel phase time-series. Then the fitted polynomial is added to the mean phase image producing a steady phase over time for each coil. Once the phase was corrected, the two series were partitioned back into the calibration and experimental series.

\begin{figure}[!h]
	\centering
	\includegraphics[width=6in]{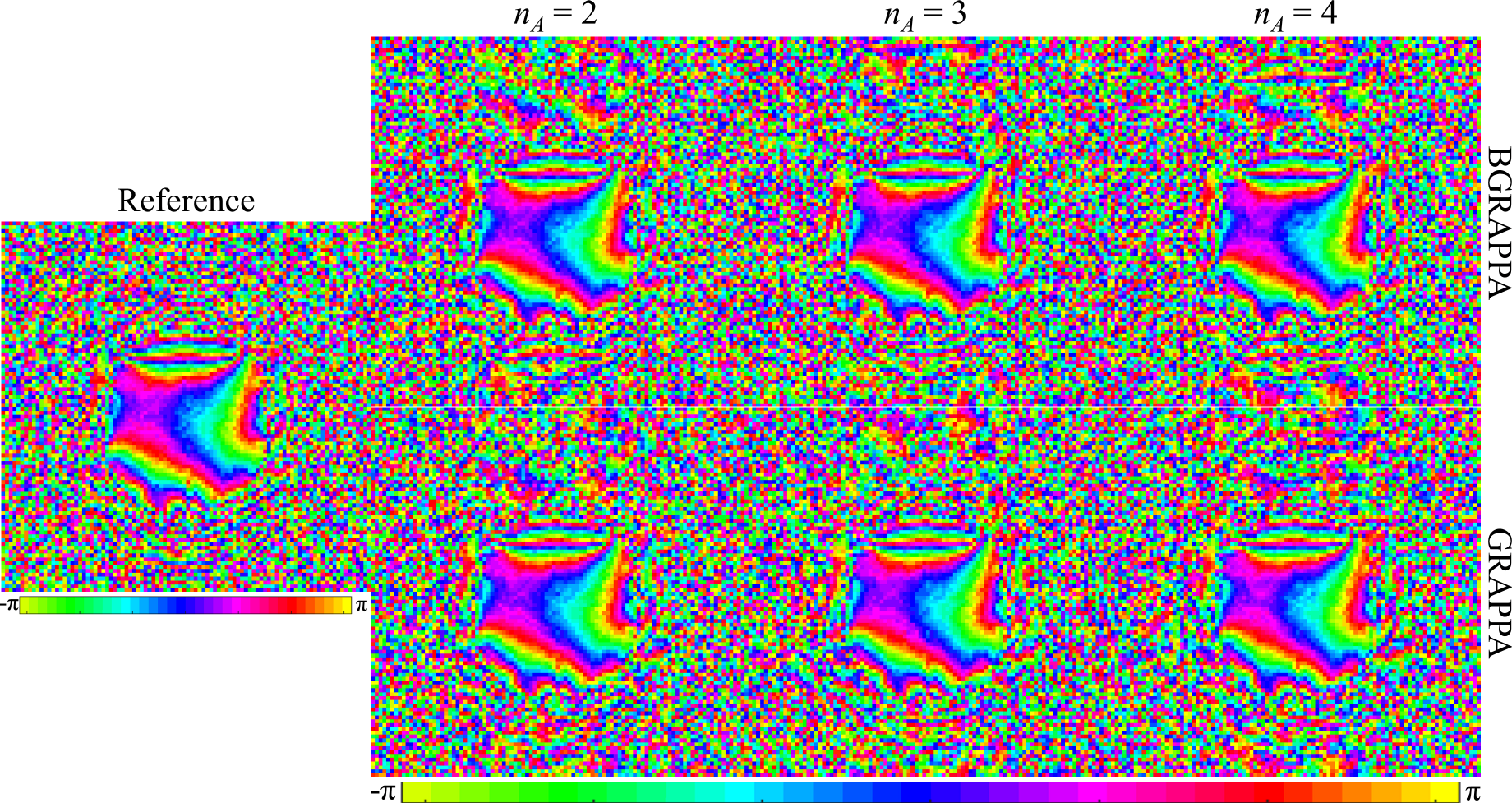}
	\caption{BGRAPPA MAP unaliased non-task phase images for each acceleration factor (first row of the right three columns) using the ICM algorithm, and GRAPPA unaliased non-task phase images for each acceleration factor (second row of the right three columns) with the phase reference image (left column). Due to the circular nature of phase angles, the color bar for the phase images have wrap-around.}
 \label{fig:ExpRecAccelTestphase}
\end{figure}

Similar to the magnitude, we also analyzed the phase of the reconstructed experimental time series using acceleration factors of ${n}_{A}=2,3,4$, displayed in Figure \ref{fig:ExpRecAccelTestphase}. The appearance of the BGRAPPA and GRAPPA phase reconstructed images are due to the imperfect shims of the magnetic field gradients. An example of this can be seen in the experimental data of the \citealt{bruce2011sense} paper. In the simulated data used for the this study, perfect homogeneity throughout the magnetic gradient field is assumed resulting in clear anatomical structure for the phase reconstructed images in Figure \ref{fig:acceltestphase} The reconstructed phase images for both methods in Figure \ref{fig:ExpRecAccelTestphase} accurately represent the experimental phase images and can also be utilized estimate the change in the $B$-field inhomogeneity, $\Delta$$B$ \citep{hahn2012magnetic}. Like the experimental magnitude images, increasing the acceleration factor has little effect of the BGRAPPA and GRAPPA reconstructed phase images.

\bibliographystyle{unsrtnat}
\bibliography{references}  




